\pdfoutput=1
\documentclass{article}

\usepackage[sort]{natbib}
\bibpunct[, ]{(}{)}{,}{a}{}{;}

\usepackage{aas_macros}

\usepackage{amsmath}
\usepackage{graphicx}
\usepackage{url}

\newcommand{\uJy}{\ensuremath{\mu{\rm Jy}}}
\newcommand{\uas}{\ensuremath{\mu{\rm as}}}
\newcommand{\degr}{\ensuremath{^\circ}}

\begin{document}

 \title{High Resolution Radio Astronomy Using Very Long Baseline
 Interferometry}
 
   \author{Enno Middelberg and Uwe Bach\thanks{{\tt middelberg@astro.rub.de, ubach@mpifr-bonn.mpg.de}}}
 
   \maketitle \begin{abstract} Very Long Baseline Interferometry, or
   VLBI, is the observing technique yielding the highest-resolution
   images today. Whilst a traditionally large fraction of VLBI
   observations is concentrating on Active Galactic Nuclei, the
   number of observations concerned with other astronomical objects
   such as stars and masers, and with astrometric applications, is
   significant. In the last decade, much progress has been made in all
   of these fields. We give a brief introduction into the technique of
   radio interferometry,  focussing on the particularities of VLBI
   observations, and review recent results which would not have been
   possible without VLBI observations.

   \end{abstract}

\section{Introduction}

Very Long Baseline Interferometry, or VLBI, is the technique by which
radio telescopes separated by up to thousands of kilometres are used
at the same time for astronomical or geodetic observations. The
signals are received and amplified at the participating antennas, are
digitized and sent to a correlator, either by storing them on tape or
disk for later shipment, or, more recently, by sending them over
network links. The correlator cross-correlates and Fourier transforms
the signals from each pair of antennas, and the result of this process
can be used to determine the brightness distribution of the sky at
radio frequencies. The angular resolution and positional accuracies
achieved in these observations are as high as a fraction of a
milli-arcsecond.

The radio regime was the first waveband accessible to astronomers
after they had studied the skies only at optical wavelengths for
hundreds of years. In the first half of the 20$^{\rm th}$ century,
rapid progress in high-frequency technology facilitated access to
radio wavelengths. Owing to the (then) relatively long wavelengths,
radio observations even with large single telescopes had poor angular
resolution of no better than 10\,arcmin, but after World War II the
first interferometers emerged and increased the resolution to less
than 1\,arcmin. Astronomers were puzzled that even at the highest
resolution achieved with radio-linked interferometers (about
0.1\,arcsec), some radio sources appeared point-like. This fuelled the
desire to use even longer baselines.

Going from directly-linked interferometers to independent elements
required recording the data in some way and processing them later, but
once the technical hurdles were overcome, the angular resolution of
these observations increased dramatically: from 1967 to 1969, the
longest baselines used in VLBI observations increased from 845\,km to
10\,592\,km, yielding an angular resolution of the order of
1\,mas. Thirty years after the first systematic radio astronomical
observations had been carried out the resolution had increased more
than a million fold. For more detailed reviews on radio interferometry
and VLBI we refer the reader to chapter one in
\cite{Thompson2001} and
\cite{Kellermann2001}.

Though many important discoveries had been made early on, VLBI
observations remain indispensable in many fields of astrophysical
research. They will much benefit from improvements in computer
technology, in particular from faster (and cheaper) network links.

This article gives a brief overview of how radio interferometry works,
of the calibration and data processing steps required to make an
image, and the various flavours and niches which can be explored. It
then reviews recent progress in the research of various astrophysical
phenomena which has been made possible with VLBI observations.

\section{The theory of interferometry and aperture synthesis}
\subsection{Fundamentals}
\label{sec:fundamentals}

\subsubsection{The visibility function}

{\it ``An interferometer is a device for measuring the spatial
coherence function''} (\citealt{Clark1999}). This dry statement pretty
much captures what interferometry is all about, and the rest of this
chapter will try to explain what lies beneath it, how the measured
spatial coherence function is turned into images and how properties of
the interferometer affect the images. We will mostly abstain from
equations here and give a written description instead, however, some
equations are inevitable. The issues explained here have been covered
in length in the literature, in particular in the first two chapters
of \cite{Taylor1999} and in great detail in \cite{Thompson2001}.

The basic idea of an interferometer is that the spatial intensity
distribution of electromagnetic radiation produced by an astronomical
object at a particular frequency, $I_\nu$, can be reconstructed from
the spatial coherence function measured at two points with the
interferometer elements, $V_\nu(\vec{r}_1, \vec{r}_2)$.

Let the (monochromatic) electromagnetic field arriving at the
observer's location $\vec{r}$ be denoted by $E_\nu(\vec{r})$. It is
the sum of all waves emitted by celestial bodies at that particular
frequency. A property of this field is the correlation function at two
points, $V_\nu(\vec{r}_1, \vec{r}_2)=\langle E_\nu(\vec{r}_1)
E^\ast_\nu(\vec{r}_2)\rangle$, where the superscript $^\ast$ denotes
the complex conjugate. $V_\nu(\vec{r}_1, \vec{r}_2)$ describes how
similar the electromagnetic field measured at two locations is. Think
of two corks thrown into a lake in windy weather. If the corks are
very close together, they will move up and down almost synchronously;
however as their separation increases their motions will become less
and less similar, until they move completely independently when
several meters apart. 

Radiation from the sky is largely spatially incoherent, except over
very small angles on the sky, and these assumptions (with a few more)
then lead to the spatial coherence function

\begin{equation}
V_\nu(\vec{r}_1, \vec{r}_2)\approx\int I_\nu(\vec{s})e^{ -2 \pi i \nu\vec{s}(\vec{r}_1-\vec{r}_2)/c}d\Omega
\end{equation}

\begin{figure}
\includegraphics[width=12cm]{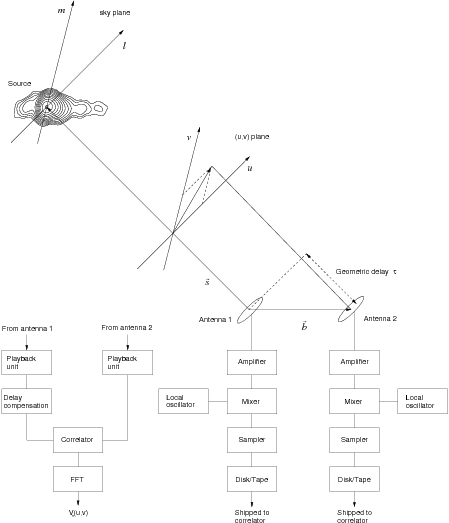}
\caption{Sketch of how a visibility measurement is obtained from a VLBI
baseline. The source is observed in direction of the line-of-sight
vector, $\vec{s}$, and the sky coordinates are the direction cosines
$l$ and $m$. The projection of the station coordinates onto the
$(u,v)$ plane, which is perpendicular to $\vec{s}$, yields the $(u,v)$
coordinates of the antennas, measured in units of the observing
wavelength. The emission from the source is delayed at one antenna by
an amount $\tau=\vec{s}\times\vec{b}/c$. At each station, the signals
are intercepted with antennas, amplified, and then mixed down to a low
frequency where they are further amplified and sampled. The essential
difference between a connected-element interferometer and a VLBI array
is that each station has an independent local oscillator, which
provides the frequency normal for the conversion from the observed
frequency to the recorded frequency. The sampled signals are written
to disk or tape and shipped to the correlator. At the correlator, the
signals are played back, the geometric delay $\tau$ is compensated
for, and the signals are correlated and Fourier transformed (in
the case of an XF correlator).}
\label{fig:schematic}
\end{figure}

Here $\vec{s}$ is the unit vector pointing towards the source and
$d\Omega$ is the surface element of the celestial sphere. The
interesting point of this equation is that it is a function of the
separation and relative orientation of two locations. An
interferometer in Europe will measure the same thing as one in
Australia, provided the separation and orientation of the
interferometer elements are the same. The relevant parameters here are
the coordinates of the antennas when projected onto a plane
perpendicular to the line of sight (Figure~\ref{fig:schematic}). This
plane has the axes $u$ and $v$, hence it is called the $(u,v)$
plane. Now let us further introduce units of wavelengths to measure
positions in the $(u,v)$ plane. One then gets

\begin{equation}
V_\nu(u,v)=\iint I_\nu(l,m)e^{-2\pi i(ul+vm)}dl\,dm
\end{equation}

This equation is a Fourier transform between the spatial coherence
function and the (modified) intensity distribution in the sky,
$I_\nu$, and can be inverted to obtain $I_\nu$. The coordinates $u$
and $v$ are the components of a vector pointing from the origin of the
$(u,v)$ plane to a point in the plane, and describe the projected
separation and orientation of the elements, measured in
wavelengths. The coordinates $l$ and $m$ are direction cosines towards
the astronomical source (or a part thereof). In radio astronomy,
$V_\nu$ is called the visibility function, but a factor, $A_\nu$, is
commonly included to describe the sensitivity of the interferometer
elements as a function of angle on the sky (the antenna
response\footnote{The fields of view in VLBI observations are
typically so small that the dependence of $(l,m)$ can be safely
ignored.  $A_\nu$ can then be set to unity and disappears.}).

\begin{equation}
V_\nu(u,v)=\iint A_\nu(l,m)I_\nu(l,m)e^{-2\pi i(ul+vm)}dl\,dm
\label{eq:visibility}
\end{equation}

The visibility function is the quantity all interferometers measure
and which is the input to all further processing by the observer.

\subsubsection{The $(u,v)$ plane}

\begin{figure}
\includegraphics[width=\linewidth]{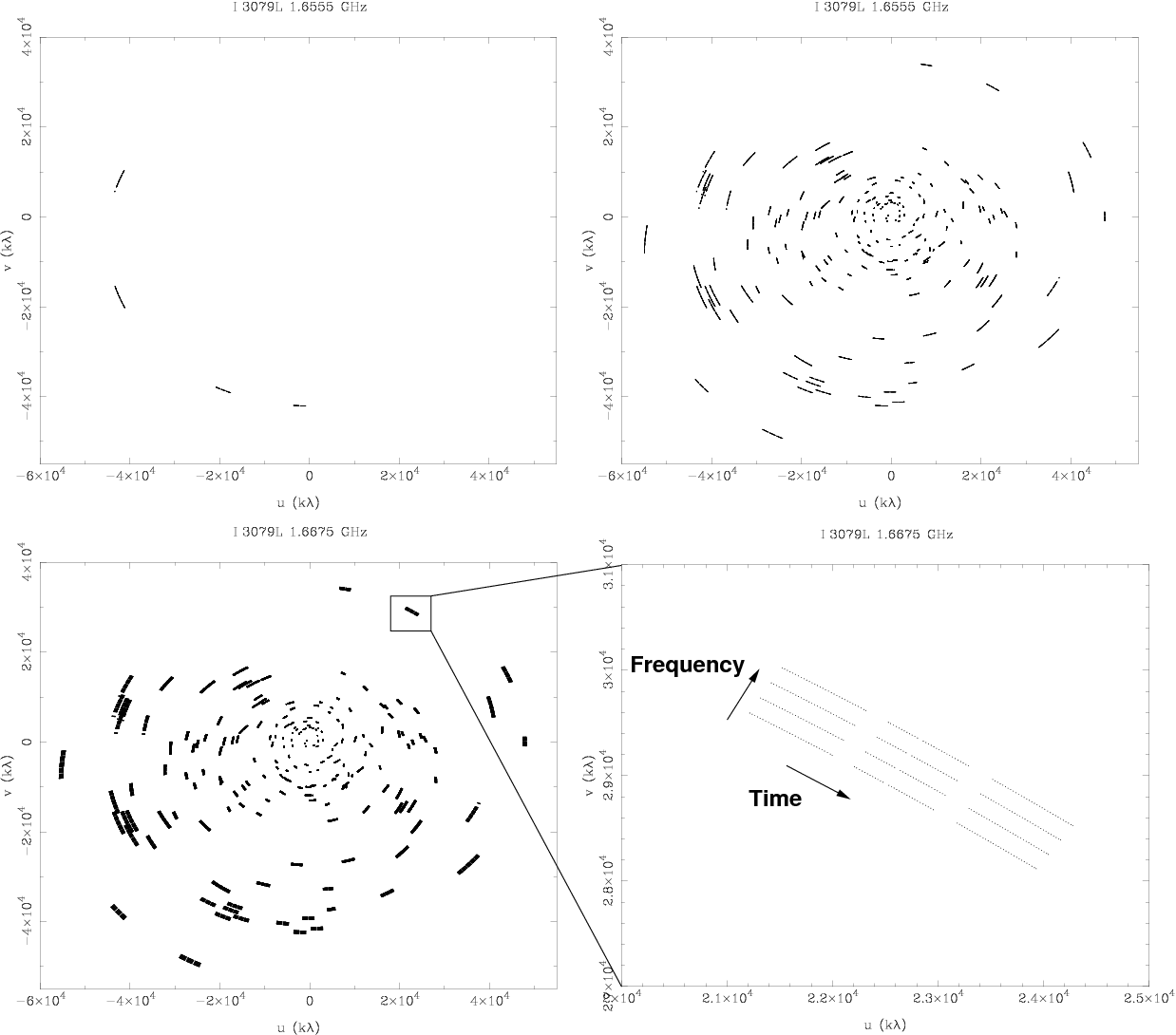}
\caption{Points in the $(u,v)$ plane sampled during a typical VLBI
observation at a frequency of 1.7\,GHz ($\lambda$=0.18\,m). The axis
units are kilolambda, and the maximum projected baseline length
corresponds to 60000\,k$\lambda$=10800\,km. The source was observed
four times for 25\,min each, over a time range of 7.5\,h. For each
visibility only one point has been plotted, i.e., the locations of the
complex conjugate visibilities are not shown. {\it Top left}: The
$(u,v)$ track of a single baseline. It describes part of an ellipse
the centre of which does not generally coincide with the origin (this
is only the case for an east-west interferometer). {\it Top right}:
The $(u,v)$ track of all interferometer pairs of all participating
antennas. {\it Bottom left}: The same as the top right panel, but the
$(u,v)$ points of all four frequency bands used in the observation
have been plotted, which broadened the tracks. {\it Bottom right}:
This plot displays a magnified portion of the previous diagram. It
shows that at any one time the four frequency bands sample different
$(u,v)$ points which lie on concentric ellipses. As the Earth rotates,
the $(u,v)$ points progress tangentially.}
\label{fig:uvplane}
\end{figure}

We introduced a coordinate system such that the line connecting the
interferometer elements, the baseline, is perpendicular to the
direction towards the source, and this plane is called the $(u,v)$
plane for obvious reasons. However, the baseline in the $(u,v)$ plane
is only a projection of the vector connecting the physical
elements. In general, the visibility function will not be the same at
different locations in the $(u,v)$ plane, an effect arising from
structure in the astronomical source. It is therefore desirable to
measure it at as many points in the $(u,v)$ plane as
possible. Fortunately, the rotation of the earth continuously changes
the relative orientation of the interferometer elements with respect
to the source, so that the point given by $(u,v)$ slowly rotates
through the plane, and so an interferometer which is fixed on the
ground samples various aspects of the astronomical source as the
observation progresses.  Almost all contemporary radio interferometers
work in this way, and the technique is then called aperture
synthesis. Furthermore, one can change the observing frequency to move
(radially) to a different point in the $(u,v)$ plane. This is
illustrated in Figure~\ref{fig:uvplane}. Note that the visibility
measured at $(-u, -v)$ is the complex conjugate of that measured at
$(u,v)$, and therefore does not add information. Hence sometimes in
plots of $(u,v)$ coverage such as Figure~\ref{fig:uvplane}, one also
plots those points mirrored across the origin. A consequence of this
relation is that after 12\,h the aperture synthesis with a given array
and frequency is complete.

\subsubsection{Image reconstruction}

After a typical VLBI observation of one source, the $(u,v)$ coverage
will not look too different from the one shown in
Figure~\ref{fig:uvplane}. These are all the data needed to form an
image by inverting Equation~\ref{eq:visibility}. However, the $(u,v)$
plane has been sampled only at relatively few points, and the purely
Fourier-transformed image (the ``dirty image'') will look poor. This
is because the true brightness distribution has been convolved with
the instrument's point-spread function (PSF). In the case of aperture
synthesis, the PSF $B(l,m)$ is the Fourier transform of the $(u,v)$
coverage:

\begin{equation}
B(l,m)=F(S(u,v))
\end{equation}

Here $S(u,v))$ is unity where measurements have been made, and zero
elsewhere. Because the $(u,v)$ coverage is mostly unsampled, $B(l,m)$
has very high artefacts (``sidelobes'').

\begin{figure}
\includegraphics[width=\linewidth]{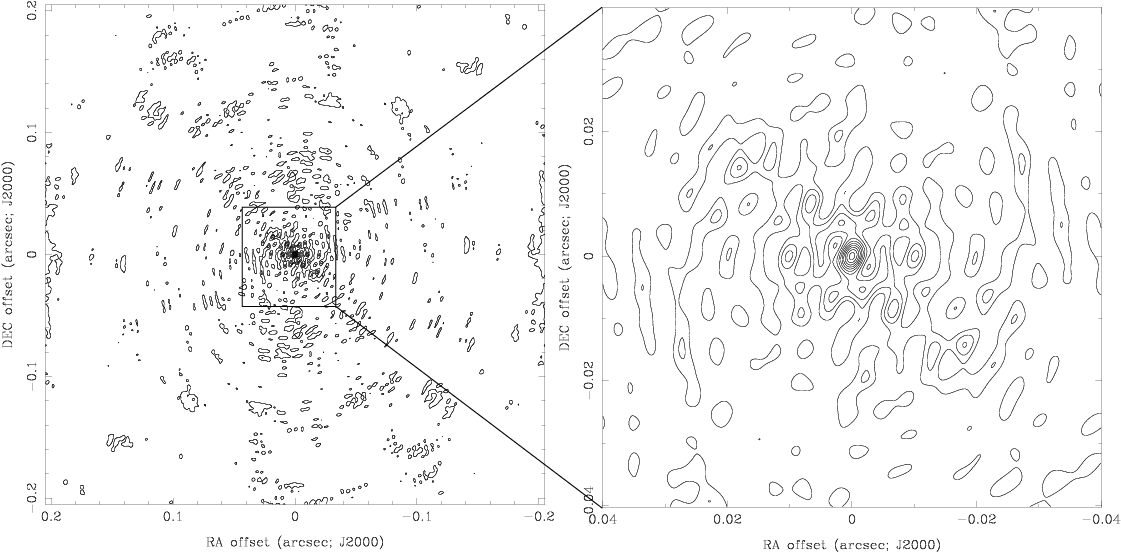}
\caption{The PSF $B(l,m)$ of the $(u,v)$ coverage shown in the bottom
left panel of Figure~\ref{fig:uvplane}. Contours are drawn at 5\,\%,
15\,\%, 25\,\%, ... of the peak response in the image centre. Patches
where the response is higher than 5\,\% are scattered all over the
image, sometimes reaching 15\,\%. In the central region, the response
outside the central peak reaches more than 25\,\%. Without further
processing, the achievable dynamic range with this sort of PSF is of
the order of a few tens.}
\label{fig:beam}
\end{figure}

To remove the sidelobes requires to interpolate the visibilities to
the empty regions of the $(u,v)$ plane, and the standard method in
radio astronomy to do that is the ``CLEAN'' algorithm.

\subsubsection{CLEAN}

The CLEAN algorithm (\citealt{Hogbom1974}) is a non-linear, iterative
mechanism to rid interferometry images of artefacts caused by
insufficient $(u,v)$ coverage. Although a few varieties exist the
basic structure of all CLEAN implementations is the same:

\begin{itemize}
\item Find the brightest pixel $I'(l_{\rm i},m_{\rm i})$ in the dirty image.

\item Translate the dirty beam $B(l,m)=F(S(u,v))$ so that
the centre pixel is at the same location as $I'(l_{\rm i},m_{\rm i})$
and scale it with a factor $c$ so that $I'(l_{\rm i},m_{\rm
i})=cB(l_{\rm i}+\Delta l,m_{\rm i}+\Delta m)$.

\item Subtract from each pixel in the dirty image the corresponding
value in the dirty beam, multiplied by a previously defined loop gain
value $\gamma<<1$ and keep the residual: $R(l,m)=I'(l,m)-c\gamma
B(l+\Delta l,m+\Delta m)$

\item Add the model component ($l_{\rm i}$, $m_{\rm i}$, $\gamma
I'(l_{\rm i},m_{\rm i})$) as a delta component to the ``clean image''
and start over again.

\end{itemize}

CLEAN can be stopped when the sidelobes of the sources in the residual
image are much lower than the image noise. A corollary of this is that
in the case of very low signal-to-noise ratios CLEAN will essentially
have no effect on the image quality, because the sidelobes are below
the noise limit introduced by the receivers in the interferometer
elements. The ``clean image'' is iteratively build up out of delta
components, and the final image is formed by convolving it with the
``clean beam''. The clean beam is a two-dimensional Gaussian which is
commonly obtained by fitting a Gaussian to the centre of the dirty
beam image. After the convolution the residual image is added.

\begin{figure}
\includegraphics[width=\linewidth]{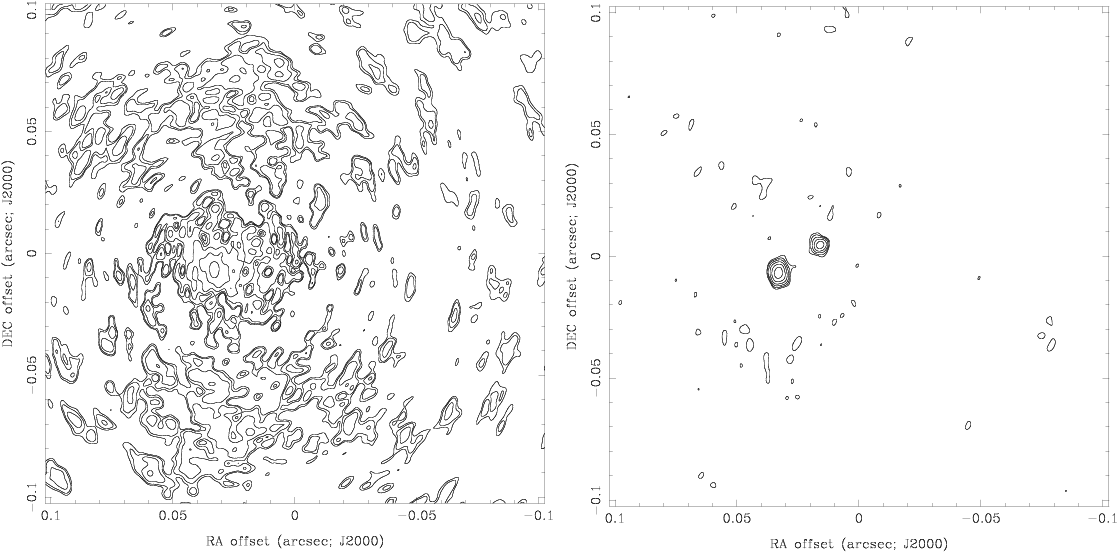}
\caption{Illustration of the effects of the CLEAN algorithm. {\it Left
panel}: The Fourier transform of the visibilities already used for
illustration in Figures~\ref{fig:uvplane} and \ref{fig:beam}. The
image is dominated by artefacts arising from the PSF of the
interferometer array. The dynamic range of the image (the image peak
divided by the rms in an empty region) is 25. {\it Right panel}: The
``clean image'', made by convolving the model components with the
clean beam, which in this case has a size of $3.0\times4.3$\,mas. The
dynamic range is 144. The contours in both panels start at 180\,\uJy\
and increase by factors of two.}
\label{fig:clean}
\end{figure}

The way in which images are formed in radio interferometry may seem
difficult and laborious (and it is), but it also adds great
flexibility. Because the image is constructed from typically thousands
of interferometer measurements one can choose to ignore measurements
from, e.g., the longest baselines to emphasize sensitivity to extended
structure. Alternatively, one can choose to weight down or ignore short
spacings to increase resolution. Or one can convolve the clean model
with a Gaussian which is much smaller than the clean beam, to make a
clean image with emphasis on fine detail (``superresolution'').

\subsubsection{Generating a visibility measurement}

The previous chapter has dealt with the fundamentals of interferometry
and image reconstruction. In this chapter we will give a brief
overview about more technical aspects of VLBI observations and the
signal processing involved to generate visibility measurements.

It may have become clear by now that an interferometer array really is
only a collection of two-element interferometers, and only at the
imaging stage is the information gathered by the telescopes
combined. In an array of $N$ antennas, the number of pairs which can
be formed is $N(N-1)/2$, and so an array of 10 antennas can measure
the visibility function at 45 locations in the $(u,v)$ plane
simultaneously. Hence, technically, obtaining visibility measurements
with a global VLBI array consisting of 16 antennas is no more complex
than doing it with a two-element interferometer - it is just
logistically more challenging.

Because VLBI observations involve telescopes at widely separated
locations (and can belong to different institutions), VLBI
observations are fully automated. The entire ``observing run'' (a
typical VLBI observations lasts around 12\,h) including setting up the
electronics, driving the antenna to the desired coordinates and
recording the raw antenna data on tape or disk, is under computer
control and requires no interaction by the observer. VLBI observations
generally are supervised by telescope operators, not astronomers.

It should be obvious that each antenna needs to point towards the
direction of the source to be observed, but as VLBI arrays are
typically spread over thousands of kilometres, a source which just
rises at one station can be in transit at another\footnote{This can be
neatly observed with the VLBA's webcam images available at
http://www.vlba.nrao.edu/sites/SITECAM/allsites.shtml. The images are
updated every 5\,min.}. Then the electronics need to be set up, which
involves a very critical step: tuning the local oscillator. In radio
astronomy, the signal received by the antenna is amplified many times
(in total the signal is amplified by factors of the order of $10^8$ to
$10^{10}$), and to avoid receiver instabilities the signal is ``mixed
down'' to much lower frequencies after the first amplification. The
mixing involves injecting a locally generated signal (the local
oscillator, or LO, signal) into the signal path with a frequency close
to the observing frequency. This yields the signal at a frequency
which is the difference between the observing frequency and the LO
frequency (see, e.g., \citealt{Rohlfs1986} for more details). The LO
frequency must be extremely stable (1 part in $10^{15}$ per day or
better) and accurately known (to the sub-Hz level) to ensure that all
antennas observe at the same frequency. Interferometers with connected
elements such as the VLA or ATCA only need to generate a single LO the
output of which can be sent to the individual stations, and any
variation in its frequency will affect all stations equally. This is
not possible in VLBI, and so each antenna is equipped with a maser
(mostly hydrogen masers) which provides a frequency standard to which
the LO is phase-locked. After downconversion, the signal is digitized
at the receiver output and either stored on tape or disk, or, more
recently, directly sent to the correlator via fast network connections
(``eVLBI'').

The correlator is sometimes referred to as the ``lens'' of VLBI
observations, because it produces the visibility measurements from the
electric fields sampled at the antennas. The data streams are aligned,
appropriate delays and phases introduced and then two operations need
to be performed on segments of the data: the cross-multiplication of
each pair of stations and a Fourier transform, to go from the temporal
domain into the spectral domain. Note that Eq.~\ref{eq:visibility} is
strictly valid only at a particular frequency. Averaging in frequency
is a source of error, and so the observing band is divided into
frequency channels to reduce averaging, and the VLBI measurand is a
cross-power spectrum.

The cross-correlation and Fourier transform can be interchanged, and
correlator designs exists which carry out the cross-correlation first
and then the Fourier transform (the ``lag-based'', or ``XF'', design
such as the MPIfR's Mark\,IV correlator and the ATCA correlator), and
also vice versa (the ``FX'' correlator such as the VLBA
correlator). The advantages and disadvantages of the two designs are
mostly in technical details and computing cost, and of little interest
to the observer once the instrument is built. However, the response to
spectral lines is different. In a lag-based correlator the signal is
Fourier transformed once after cross-correlation, and the resulting
cross-power spectrum is the intrinsic spectrum convolved with the sinc
function. In a FX correlator, the input streams of both interferometer
elements are Fourier transformed which includes a convolution with the
sinc function, and the subsequent cross-correlation produces a
spectrum which is convolved with the square of the sinc
function. Hence the FX correlator has a finer spectral resolution and
lower sidelobes (see \citealt{Romney1999}).

The vast amount of data which needs to be processed in interferometry
observations has always been processed on purpose-built computers
(except for the very observations where bandwidths of the order of
several hundred kHz were processed on general purpose computers). Only
recently has the power of off-the-shelf PCs reached a level which
makes it feasible to carry out the correlation in
software. \cite{Deller2007} describe a software correlator which can
efficiently run on a cluster of PC-architecture computers. Correlation
is an ``embarrassingly parallel'' problem, which can be split up in
time, frequency, and by baseline, and hence is ideally suited to run
in a cluster environment.

The result of the correlation stage is a set of visibility
measurements. These are stored in various formats along with auxiliary
information about the array such as receiver temperatures, weather
information, pointing accuracy, and so forth. These data are sent to
the observer who has to carry out a number of calibration steps before
the desired information can be extracted.

\subsection{Sources of error in VLBI observations}
\label{sec:errors}

VLBI observations generally are subject to the same problems as
observations with connected-element interferometers, but the fact that
the individual elements are separated by hundreds and thousands of
kilometres adds a few complications.

The largest source of error in typical VLBI observations are phase
errors introduced by the earth's atmosphere. Variations in the
atmosphere's electric properties cause varying delays of the radio
waves as they travel through it. The consequence of phase errors is
that the measured flux of individual visibilities will be scattered
away from the correct locations in the image, reducing the SNR of the
observations or, in fact, prohibiting a detection at all.  Phase
errors arise from tiny variations in the electric path lengths from
the source to the antennas. The bulk of the ionospheric and
tropospheric delays is compensated in correlation using atmospheric
models, but the atmosphere varies on very short timescales so that
there are continuous fluctuations in the measured visibilities.

At frequencies below 5\,GHz changes in the electron content (TEC) of
the ionosphere along the line of sight are the dominant source of
phase errors. The ionosphere is the uppermost part of the earth's
atmosphere which is ionised by the sun, and hence undergoes diurnal
and seasonal changes. At low GHz frequencies the ionosphere's plasma
frequency is sufficiently close to the observing frequency to have a
noticeable effect. Unlike tropospheric and most other errors, which
have a linear dependence on frequency, the impact of the ionosphere is
proportional to the inverse of the frequency squared, and so fades
away rather quickly as one goes to higher frequencies. Whilst the TEC
is regularly monitored\footnote{http://iono.jpl.nasa.gov} and the
measurements can be incorporated into the VLBI data calibration, the
residual errors are still considerable.

At higher frequencies changes in the tropospheric water vapour content
have the largest impact on radio interferometry observations. Water
vapour\footnote{note that clouds do not primarily consist of water
vapour, but of condensated water in droplets.} does not mix well with
air and thus the integrated amount of water vapour along the line of
sight varies considerably as the wind blows over the
antenna. Measuring the amount of water vapour along the line of sight
is possible and has been implemented at a few observatories
(Effelsberg, CARMA, Plateau de Bure), however it is difficult and not
yet regularly used in the majority of VLBI observations.

Compact arrays generally suffer less from atmospheric effects because
most of the weather is common to all antennas. The closer two antennas
are together, the more similar the atmosphere is along the lines of
sight, and the delay difference between the antennas decreases.

Other sources of error in VLBI observations are mainly uncertainties
in the geometry of the array and instrumental errors. The properties
of the array must be accurately known in correlation to introduce the
correct delays. As one tries to measure the phase of an
electromagnetic wave with a wavelength of a few centimetres, the array
geometry must be known to a fraction of that. And because the earth is
by no means a solid body, many effects have to be taken into account,
from large effects like precession and nutation to smaller effects
such as tectonic plate motion, post-glacial rebound and gravitational
delay. For an interesting and still controversial astrophysical
application of the latter, see \cite{Fomalont2003}. For a long list of
these effects including their magnitudes and timescales of
variability, see \cite{Walker1999}.

\subsection{The problem of phase calibration: self-calibration}

Due to the aforementioned errors, VLBI visibilities directly from the
correlator will never be used to make an image of astronomical
sources. The visibility phases need to be calibrated in order to
recover the information about the source's location and
structure. However, how does one separate the unknown source
properties from the unknown errors introduced by the instrument and
atmosphere? The method commonly used to do this is called
self-calibration and works as follows.

In simple words, in self-calibration one uses a model of the source
(if a model is not available a point source is used) and tries to find
phase corrections for the antennas to make the visibilities comply with
that model. This won't work perfectly unless the model was a perfect
representation of the source, and there will be residual, albeit
smaller, phase errors. However the corrected visibilities will allow
one to make an improved source model, and one can find phase
corrections to make the visibilities comply with that improved model,
which one then uses to make an even better source model. This process
is continued until convergence is reached.

The assumption behind self-calibration is that the errors in the
visibility phases, which are baseline-based quantities, can be
described as the result of antenna-based errors. Most of the errors
described in Sec~\ref{sec:errors} are antenna-based: e.g. delays
introduced by the atmosphere, uncertainties in the location of the
antennas, and drifts in the electronics all are antenna-based. The
phase error of a visibility is the combination of the antenna-based
phase errors\footnote{Baseline-based errors exist, too, but are far
less important, see \cite{Cornwell1999} for a list.}. Since the number
of unknown station phase errors is less than the number of
visibilities, there is additional phase information which can be used
to determine the source structure. 

However, self-calibration contains some traps. The most important is
making a model of the source which is usually accomplished by making a
deconvolved image with the CLEAN algorithm. If a source has been
observed during an earlier epoch and the structural changes are
expected to be small, then one can use an existing model for a
start. If in making that model one includes components which are not
real (e.g., by cleaning regions of the image which in fact do not
contain emission) then they will be included in the next iteration of
self-calibration and will re-appear in the next image. Although it is
not easy to generate fake sources or source parts which are strong,
weaker source structures are easily affected. The authors have
witnessed a radio astronomy PhD student producing a map of a
colleague's name using a data set of pure noise, although the SNR was
of the order of only 3.

It is also important to note that for self-calibration the SNR of the
visibilities needs to be of the order of 5 or higher within the time
it takes the fastest error component to change by a few tens of
degrees (``atmospheric coherence time'') (\citealt{Cotton1995b}). Thus
the integration time for self-calibration usually is limited by
fluctuations in the tropospheric water vapour content. At 5\,GHz, one
may be able to integrate for 2\,min without the atmosphere changing
too much, but this can drop to as little as 30\,s at 43\,GHz. Because
radio antennas used for VLBI are less sensitive at higher frequencies
observations at tens of GHz require brighter and brighter sources to
calibrate the visibility phases. Weak sources well below the detection
threshold within the atmospheric coherence time can only be observed
using phase referencing (see Sec.~\ref{sec:phase-referencing}.

Another boundary condition for successfully self-calibration is that
for a given number of array elements the source must not be too
complex. The more antennas, the better because the ratio of number of
constraints to number of antenna gains to be determined goes as $N/2$
(The number of constraints is the number of visibilities,
$N(N-1)/2$. The number of gains is the number of stations, $N$, minus
one, because the phase of one station is a free parameter and set to
zero). Thus self-calibration works very well at the VLA, even with
complex sources, whereas for an east-west interferometer with few
elements such as the ATCA, self-calibration is rather limited. In VLBI
observations, however, the sources are typically simple enough to make
self-calibration work even with a modest number ($N>5$) of antennas.

\subsubsection{Phase referencing}
\label{sec:phase-referencing}

It is possible to obtain phase-calibrated visibilities without
self-calibration by measuring the phase of a nearby, known
calibrator. The assumption is that all errors for the calibrator and
the target are sufficiently similar to allow calibration of the target
with phase corrections derived from the calibrator. While this
assumption is justified for the more slowly varying errors such as
clock errors and array geometry errors (provided target and calibrator
are close), it is only valid under certain circumstances when
atmospheric errors are considered. The critical ingredients in phase
referencing observations are the target-calibrator separation and the
atmospheric coherence time. The separation within which the phase
errors for the target and calibrator are considered to be the same is
called the isoplanatic patch, and is of the order of a few degrees at
5\,GHz. The switching time must be shorter than the atmospheric
coherence time to prevent undersampling of the atmospheric phase
variations. At high GHz frequencies this can result in observing
strategies where one spends half the observing time on the calibrator.

Phase-referencing not only allows one to observe sources too weak for
self-calibration, but it also yields precise astrometry for the target
relative to the calibrator. A treatment of the attainable accuracy can
be found in \cite{Pradel2006}.

\subsection{Polarization}

The polarization of radio emission can yield insights into the
strength and orientation of magnetic fields in astrophysical objects
and the associated foregrounds. As a consequence and because the
calibration has become easier and more streamlined it has become
increasingly popular in the past 10 years to measure polarization.

Most radio antennas can record two orthogonal polarizations,
conventionally in the form of dual circular polarization. In
correlation, one can correlate the right-hand circular polarization
signal (RCP) of one antenna with the left-hand circular polarization
(LCP) of another and vice versa, to obtain the cross-polarization
products $RL$ and $LR$. The parallel-hand circular polarization
cross-products are abbreviated as $RR$ and $LL$. The four correlation
products are converted into the four Stokes parameters in the
following way:

\begin{equation}
\begin{split}
I & =  \frac{1}{2} (RR + LL)\\
Q & =  \frac{1}{2} (RL + LR)\\
U & = j\frac{1}{2} (LR - RL)\\
V & =  \frac{1}{2} (RR - LL).\\
\end{split}
\end{equation}

From the Stokes images one can compute images of polarized intensity
and polarization angle.

Most of the calibration in polarization VLBI observations is identical
to conventional observations, where one either records only data in
one circular polarization or does not form the cross-polarization data
at the correlation stage. However, two effects need to be taken care
of, the relative phase relation between RCP and LCP and the leakage of
emission from RCP and LCP into the cross-products.

The relative phase orientation of RCP and LCP needs to be calibrated
to obtain the absolute value for the electric vector position angle
(EVPA) of the polarized emission in the source. This is usually
accomplished by observing a calibrator which is known to have a stable
EVPA with a low-resolution instrument such as a single dish telescope
or a compact array.

Calibration of the leakage is more challenging. Each radio telescope
has polarization impurities arising from structural asymmetries and
errors in manufacturing, resulting in ``leakage'' of emission from one
polarization to the other. The amount of leakage typically is of the
order of a few percent and thus is of the same order as the typical
degree of polarization in the observed sources and so needs to be
carefully calibrated. The leakage is a function of frequency but can
be regarded as stable over the course of a VLBI observation.

Unfortunately, sources which are detectable with VLBI are extremely
small and hence mostly variable. It is therefore not possible to
calibrate the leakage by simply observing a polarization calibrator,
and the leakage needs to be calibrated by every observer. At present
the calibration scheme exploits the fact that the polarized emission
arising from leakage does not change its position angle in the course
of the observations. The EVPA of the polarized emission coming from
the source, however, will change with respect to the antenna and its
feed horns, because most antennas have alt-azimuth mounts and so the
source seems to rotate on the sky as the observation
progresses\footnote{The 26\,m antenna at the Mount Pleasant
Observatory near Hobart, Australia, has a parallactic mount and thus
there is no field rotation.}. One can think of this situation as the
sum of two vectors, where the instrumental polarization is a fixed
vector and the astronomical polarization is added to this vector and
rotates during the observation. Leakage calibration is about
separating these two contributions, by observing a strong source at a
wide range of position angles. The method is described in
\cite{Leppanen1995}, and a more detailed treatment of polarization
VLBI is given by
\cite{Kemball1999}.

\subsection{Spectral line VLBI}

In general a set of visibility measurements consists of cross-power
spectra. If a continuum source has been targeted, the number of
spectral points is commonly of the order of a few tens. If a spectral
line has been observed, the number of channels can be as high as a few
thousand, and is limited by the capabilities of the correlator. The
high brightness temperatures (Section~\ref{sec:tb}) needed to yield a
VLBI detection restrict observations to masers, or relatively large
absorbers in front of non-thermal continuum sources. The setup of
frequencies requires the same care as for short baseline
interferometry, but an additional complication is that the antennas
have significant differences in their Doppler shifts towards the
source. See \cite{Westpfahl1999}, \cite{Rupen1999}, and \cite{Reid1999}
for a detailed treatment of spectral-line interferometry and VLBI.

\subsection{Pulsar gating}

If pulsars are to be observed with a radio interferometer it is
desirable to correlate only those times where a pulse is arriving (or
where it is absent, \citealt{Stappers1999}). This is called pulsar
gating and is an observing mode available at most interferometers.

\subsection{Wide-field limitations}

The equations in Sec~\ref{sec:fundamentals} are strictly correct only
for a single frequency and single points in time, but radio telescopes
must observe a finite bandwidth, and in correlation a finite
integration time must be used, to be able to detect objects. Hence a
point in the $(u,v)$ plane always represents the result of averaging
across a bandwidth, $\Delta\nu$, and over a time interval, $\Delta\tau$
(the {\it points} in Fig.~\ref{fig:uvplane} actually represent a
continuous frequency band in the radial direction and a continuous
observation in the time direction).

The errors arising from averaging across the bandwidth are referred to
as bandwidth smearing, because the effect is similar to chromatic
aberration in optical systems, where the light from one single point
of an object is distributed radially in the image plane. In radio
interferometry, the images of sources away from the observing centre
are smeared out in the radial direction, reducing the signal-to-noise
ratio. The effect of bandwidth smearing increases with the fractional
bandwidth, $\Delta\nu/\nu$, the square root of the distance to the
observing centre, $\sqrt{l^2+m^2}$, and with $1/\theta_b$, where
$\theta_b$ is the FWHM of the synthesized beam. Interestingly,
however, the dependencies of $\nu$ of the fractional bandwidth and of
$\theta_b$ cancel one another, and so for any given array and
bandwidth, bandwidth smearing is independent of $\nu$ (see
\citealt{Thompson2001}). The effect can be avoided if the observing
band is subdivided into a sufficiently large number of frequency
channels for all of which one calculates the locations in the $(u,v)$
plane separately. This technique is sometimes deliberately chosen to
increase the $(u,v)$ coverage, predominantly at low frequencies where
the fractional bandwidth is large. It is then called multi-frequency
synthesis.

By analogy, the errors arising from time averaging are called time
smearing, and they smear out the images approximately tangentially to
the $(u,v)$ ellipse. It occurs because each point in the $(u,v)$ plane
represents a measurement during which the baseline vector rotated
through $\omega_E\Delta\tau$, where $\omega_E$ is the angular velocity
of the earth. Time smearing also increases as a function of
$\sqrt{l^2+m^2}$ and can be avoided if $\Delta\tau$ is chosen small
enough for the desired field of view.

VLBI observers generally are concerned with fields of view (FOV) of no
more than about one arcsecond, and consequently most VLBI observers
are not particularly bothered by wide field effects. However,
wide-field VLBI has gained momentum in the last few years as the
computing power to process finely time- and bandwidth-sampled data sets
has become widely available. Recent examples of observations with
fields of view of 1$^\prime$ or more are reported on in
\cite{McDonald2001},
\cite{Garrett2001}, \cite{Garrett2005}, \cite{Lenc2006} and
\cite{Lenc2006b}. The effects of primary beam attenuation, bandwidth
smearing and time smearing on the SNR of the observations can be
estimated using the calculator at
http://astronomy.swin.edu.au/\~{}elenc/Calculators/wfcalc.php.

\subsection{VLBI at mm wavelengths}
\label{sec:mm-vlbi}

In the quest for angular resolution VLBI helps to optimize one part of
the equation which approximates the separation of the finest details
an instrument is capable of resolving, $\theta=\lambda/D$ . In VLBI,
$D$ approaches the diameter of the earth, and larger instruments are
barely possible, although ``space-VLBI'' can extend an array beyond
earth. However, it is straightforward to attempt to decrease $\lambda$
to push the resolution further up.

However, VLBI observations at frequencies above 20\,GHz
($\lambda=$15\,mm) become progressively harder towards higher
frequencies. Many effects contribute to the difficulties at mm
wavelengths: the atmospheric coherence time is shorter than one
minute, telescopes are less efficient, receivers are less sensitive,
sources are weaker than at cm wavelengths, and tropospheric water
vapour absorbs the radio emission. All of these effects limit mm VLBI
observations to comparatively few bright continuum sources or sources
hosting strong masers. Hence also the number of possible phase
calibrators for phase referencing drops. Nevertheless VLBI
observations at 22\,GHz (13\,mm) , 43\,GHz ($\lambda\approx$7\,mm) and
86\,GHz ($\lambda\approx$3\,mm) are routinely carried out with the
world's VLBI arrays. For example, of all projects observed in 2005 and
2006 with the VLBA, 16\,\% were made at 22\,GHz, 23\,\% at 43\,GHz,
and 4\,\% at 86\,GHz\footnote{e.g.,
ftp://ftp.aoc.nrao.edu/pub/cumvlbaobs.txt}.

Although observations at higher frequencies are experimental, a
convincing demonstration of the feasibility of VLBI at wavelengths
shorter than 3\,mm was made at 2\,mm (147GHz) in 2001 and 2002
\citep{2002evlb.conf..125K,2002A&A...390L..19G}. These first 2\,mm-VLBI experiments
resulted in detections of about one dozen quasars on the short
continental and long transatlantic baselines
\citep{2002evlb.conf..125K}. In an experiment in April 2003 at 1.3\,mm
(230GHz) a number of sources was detected on the 1150\,km long
baseline between Pico Veleta and Plateau de Bure
\citep{2004evn..conf...15K}. On the 6.4\,G $\lambda$ long
transatlantic baseline between Europe and Arizona, USA fringes for the
quasar 3C454.3 were clearly seen. This detection marks a new record in
angular resolution in Astronomy (size $<30\,\mu$as). It indicates the
existence of ultra compact emission regions in AGN even at the highest
frequencies (for 3C454.3 at z=0.859, the rest frame frequency is
428\,GHz). So far, no evidence for a reduced brightness temperature of
the VLBI-cores at mm wavelengths was found
\citep{2004evn..conf...15K}. These are the astronomical observations
with the highest angular resolution possible today at any wavelength.

\subsection{The future of VLBI: eVLBI, VLBI in space, and the SKA}

One of the key drawbacks of VLBI observations has always been that the
raw antenna signals are recorded and the visibilities formed only
later when the signals are combined in the correlator. Thus there has
never been an immediate feedback for the observer, who has had to wait
several weeks or months until the data are received for
investigation. With the advent of fast computer networks this has
changed in the last few years. First small pieces of raw data were
sent from the antennas to the correlator, to check the integrity of
the VLBI array, then data were directly streamed onto disks at the
correlator, then visibilities were produced in real-time from the data
sent from the antennas to the correlator. A brief description of the
transition from tape-based recording to real-time correlation of the
European VLBI Network (EVN) is given in
\cite{Szomoru2004}. The EVN now regularly performs so-called ``eVLBI''
observing runs which is likely to be the standard mode of operation in
the near future. The Australian Long Baseline Array (LBA) completed a
first eVLBI-only observing run in March
2007\footnote{http://wwwatnf.atnf.csiro.au/vlbi/evlbi}.

It has been indicated in Sec.~\ref{sec:mm-vlbi} that resolution can be
increased not only by observing at higher frequencies with
ground-based arrays but also by using a radio antenna in earth
orbit. This has indeed been accomplished with the Japanese satellite
``HALCA'' (Highly Advanced Laboratory for Communications and
Astronomy, \citealt{Hirabayashi2000}) which was used for VLBI
observations at 1.6\,GHz, 5\,GHz and 22\,GHz (albeit with very low
sensitivity) between 1997 and 2003. The satellite provided the
collecting area of an 8\,m radio telescope and the sampled signals
were directly transmitted to ground-based tracking stations. The
satellite's elliptical orbit provided baselines between a few hundred
km and more than 20\,000\,km, yielding a resolution of up to 0.3\,mas
(\citealt{Dodson2006,Edwards2002}). Amazingly, although HALCA only
provided left-hand circular polarization, it has been used
successfully to observe polarized emission (e.g.,
\citealt{Kemball2000,2006A&A...452...83B}). But this was only possible
because the ground array observed dual circular polarization. Many of
the scientific results from VSOP are summarized in two special issues
of Publications of the Astronomical Society of Japan (PASJ, Vol.\ 52,
No.\ 6, 2000 and Vol.\ 58, No.\ 2, 2006). The successor to HALCA,
ASTRO-G, is under development and due for launch in 2012. It will have
a reflector with a diameter of 9\,m and receivers for observations at
8\,GHz, 22\,GHz, and 43\,GHz.

The Russian mission RadioAstron is a similar project to launch a 10\,m
radio telescope into a high apogee orbit. It will carry receivers for
frequencies between 327\,MHz and 25\,GHz, and is due for launch in
October 2008.

The design goals for the Square Kilometre Array (SKA), a large,
next-generation radio telescope built by an international consortium,
include interferometer baselines of at least 3000\,km. At the same
time, the design envisions the highest observing frequency to be
25\,GHz, and so one would expect a maximum resolution of around
1\,mas. However, most of the baselines will be shorter than 3000\,km,
and so a weighted average of all visibilities will yield a resolution
of a few mas, and of tens of mas at low GHz frequencies. The SKA's
resolution will therefore be comparable to current VLBI arrays. Its
sensitivity, however, will be orders of magnitude higher (sub-\uJy\ in
1\,h). The most compelling consequence of this is that the SKA will
allow one to observe thermal sources with brightness temperatures of
the order of a few hundred Kelvin with a resolution of a few
mas. Current VLBI observations are limited to sources with brightness
temperatures of the order of $10^6$\,K and so to non-thermal radio
sources and coherent emission from masers. With the SKA one can
observe star and black hole formation throughout the universe, stars,
water masers at significant redshifts, and much more. Whether or not
the SKA can be called a VLBI array in the original sense (an array of
isolated antennas the visibilities of which are produced later on) is
a matter of taste: correlation will be done in real time and the local
oscillator signals will be distributed from the same source. Still,
the baselines will be ``very long'' when compared to 20th century
connected-element interferometers. A short treatment of ``VLBI with
the SKA'' can be found in
\cite{Carilli2005}. Comprehensive information about the current state
of the SKA is available on the aforementioned web page; prospects of
the scientific outcomes of the SKA are summarized in
\cite{Carilli2004}; and engineering aspects are treated in
\cite{Hall2005}.

\subsection{VLBI arrays around the world and their capabilities}

This section gives an overview of presently active VLBI arrays which
are available for all astronomers and which are predominantly used for
astronomical observations. Antennas of these arrays are frequently
used in other array's observations, to either add more long baselines
or to increase the sensitivity of the observations. Joint observations
including the VLBA and EVN antennas are quite common; also
observations with the VLBA plus two or more of the phased VLA, the
Green Bank, Effelsberg and Arecibo telescopes (then known as the High
Sensitivity Array) have recently been made easier through a common
application process. Note that each of these four telescopes has more
collecting area than the VLBA alone, and hence the sensitivity
improvement is considerable.

\begin{table}[htbp]
\caption[Properties of AGN.]{VLBI telescope information on the World Wide Web.}
\label{tab:telescopes}
\centering
\begin{tabular}{l|l}
\hline\hline
Square Kilometre Array 	 & http://www.skatelescope.org \\
High Sensitivity Array 	 & http://www.nrao.edu/HSA \\
European VLBI Network  	 & http://www.evlbi.org \\
Very Long Baseline Array & http://www.vlba.nrao.edu \\
Long Baseline Array      & http://www.atnf.csiro.au/vlbi \\
VERA                     & http://veraserver.mtk.nao.ac.jp/outline/index-e.html \\
GMVA                     & http://www.mpifr-bonn.mpg.de/div/vlbi/globalmm\\
\hline
\end{tabular}
\end{table}

\subsubsection{The European VLBI Network (EVN)}

The EVN is a collaboration of 14 institutes in Europe, Asia, and South
Africa and was founded in 1980. The participating telescopes are used
in many independent radio astronomical observations, but are scheduled
three times per year for several weeks together as a VLBI array. The
EVN provides frequencies in the range of 300\,MHz to 43\,GHz, though
due to its inhomogeneity not all frequencies can be observed at all
antennas. The advantage of the EVN is that it includes several
relatively large telescopes such as the 76\,m Lovell telescope, the
Westerbork array, and the Effelsberg 100\,m telescope, which provide
high sensitivity. Its disadvantages are a relatively poor frequency
agility during the observations, because not all telescopes can change
their receivers at the flick of a switch. EVN observations are mostly
correlated on the correlator at the Joint Institute for VLBI in Europe
(JIVE) in Dwingeloo, the Netherlands, but sometimes are processed at
the Max-Planck-Institute for Radio Astronomy in Bonn, Germany, or the
National Radio Astronomy Observatory in Socorro, USA.

\subsubsection{The U.S. Very Long Baseline Array (VLBA)}

The VLBA is a purpose-built VLBI array across the continental USA and
islands in the Caribbean and Hawaii. It consists of 10 identical
antennas with a diameter of 25\,m, which are remotely operated from
Socorro, New Mexico. The VLBA was constructed in the early 1990s and
began full operations in 1993. It provides frequencies between
300\,MHz and 86\,GHz at all stations (except two which are not worth
equipping with 86\,GHz receivers due to their humid locations). Its
advantages are excellent frequency agility and its homogeneity, which
makes it very easy to use. Its disadvantages are its comparatively
small antennas, although the VLBA is frequently used in conjunction
with the phased VLA and the Effelsberg and Green Bank telescopes.

\subsubsection{The Australian Long Baseline Array (LBA)}

The LBA consists of six antennas in Ceduna, Hobart, Parkes, Mopra,
Narrabri, and Tidbinbilla. Like the EVN it has been formed from
existing antennas and so the array is inhomogeneous. Its frequency
range is 1.4\,GHz to 22\,GHz, but not all antennas can observe at all
available frequencies. Stretched out along Australia's east coast, the
LBA extends in a north-south direction which limits the $(u,v)$
coverage. Nevertheless, the LBA is the only VLBI array which can
observe the entire southern sky, and the recent technical developments
are remarkable: the LBA is at the forefront of development of eVLBI
and at present is the only VLBI array correlating all of its
observations using the software correlator developed by
\cite{Deller2007} on a computer cluster of the Swinburne Centre for
Astrophysics and Supercomputing.

\subsubsection{The Korean VLBI Network (KVN)}

The KVN is a dedicated VLBI network consisting of three antennas which
currently is under construction in Korea. It will be able to observe
at up to four widely separated frequencies (22\,GHz, 43\,GHz, 86\,GHz,
and 129\,GHz), but will also be able to observe at 2.3\,GHz and
8.4\,GHz. The KVN has been designed to observe H$_2$O and SiO masers
in particular and can observe these transitions
simultaneously. Furthermore, the antennas can slew quickly for
improved performance in phase-referencing observations.

\subsubsection{The Japanese VERA network}

VERA (VLBI Exploration of Radio Astrometry) is a purpose-built VLBI
network of four antennas in Japan. The scientific goal is to measure
the annual parallax towards galactic masers (H$_2$O masers at 22\,GHz
and SiO masers at 43\,GHz), to construct a map of the Milky
Way. Nevertheless, VERA is open for access to carry out any other
observations. VERA can observe two sources separated by up to
2.2\degr\ simultaneously, intended for an extragalactic reference
source and for the galactic target. This observing mode is a
significant improvement over the technique of phase-referencing where
the reference source and target are observed in turn. The positional
accuracy is expected to reach 10\,$\mu$as, and recent results seem to
reach this (\citealt{Hirota2007}). VERA's frequency range is 2.3\,GHz
to 43\,GHz.

\subsubsection{The Global mm-VLBI Array (GMVA)}

The GMVA is an inhomogeneous array of 13 radio telescopes capable of
observing at a frequency of 86\,GHz. Observations with this network
are scheduled twice a year for about a week. The array's objective is
to provide the highest angular resolution on a regular schedule.

\section{Astrophysical applications}
High resolution radio observations are used in many astronomical
fields. They provide information about the structures around Active
Galactic Nuclei (AGN), their accretion disks, powerful relativistic
jets, magnetic fields, and absorbing material, and they are used to
image peculiar binary stars, radio-stars and young supernova remnants.
Spectroscopic VLBI-observations allow one to investigate compact maser
emitting regions of various kinds in galaxies, star birth regions and
stellar envelopes. The phase-referencing technique allows to determine
source positions with highest possible accuracy. This section will
summarize some of the observational highlights obtained by very long
baseline radio interferometers.

\subsection{Active Galactic Nuclei and their jets}

The luminosity of most of the known galaxies is dominated in the
optical by their stellar emission, but in some galaxies a significant
fraction of the energy output has a non-thermal origin. These are
called active galaxies or Active Galactic Nuclei (AGN), if the
non-thermal emission comes mainly from the core.  Although AGN
represent only a small fraction of all galaxies they have been studied
intensively over the last 40 years at all accessible wavelengths. This
is partly because their high luminosities made them the only objects
that could be studied at cosmologically significant distances for a
long time. At radio wavelength some active galaxies are the most
luminous sources in the sky and were the first sources detected in the
early days of radio astronomy (e.g. \citealt{1954ApJ...119..206B}).

In many powerful radio galaxies, the radio emission is detected
several hundreds of kiloparsec away from the centre of the associated
optical galaxy \citep[e.g.][and references
therein]{1984ARA&A..22..319B}. These regions are called radio
lobes. In other cases, however, a large portion of the radio emission
comes from much smaller regions at the centre of the galaxy with
dimensions of only a few parsecs. These centres are also visible in
lobe-dominated objects, and they supply the lobes with plasma via
relativistic outflows, which are called jets.

Since single-dish radio telescopes have rather low resolution,
observations of jets and lobes in radio galaxies were one of the main
drivers for the development of radio interferometers in the early
1960s (\citealt{1960MNRAS.120..220R}). In particular the development
of Very Long Baseline Interferometry (VLBI) made it possible to image
the inner regions of AGN and it is still the only technique capable to
image sub-parsec-scale structures in extra-galactic objects. The
direct imaging of parsec-scale jets and the complementary study of
activity in the associated AGN in all spectral regimes has broadly
enhanced our understanding of these objects. Based on the last decades
of detailed observations of AGN a common scheme about their nature has
been established.

\subsubsection{The AGN paradigm}

It is believed that the high luminosities of AGN are due to accretion
on to a super-massive black hole
\citep[e.g.][]{1984RvMP...56..255B,2001Sci...291...84M}.  The
10$^{6-10}\,M_\odot$ black hole is surrounded by an accretion disc and
probably a hot corona mainly radiating at optical to soft X-ray
energies \citep{1985ApJ...297..621A}. In the vicinity of the nucleus
high velocity gas, characterized by broad optical emission lines,
forms the so called broad-line region (BLR), and lower velocity gas
with narrower emission lines forms the narrow-line region (NLR). An
obscuring torus of gas and dust is hiding the central region including
the BLR from some directions. The term ``torus'' is often used in this
context, but it is not sure at all which geometrical form the absorber
has. At the poles relativistic jets are formed and expand for several
tens of kiloparsec or even up to megaparsecs
\citep[e.g.][]{1995PASP..107..803U}.

Not all AGN show observational signatures of every
component. Therefore several classes were defined that divide AGN
mainly into type 1 source which exhibit broad emission lines from the
BLR and type 2 sources that do not show broad emission lines. Both
groups contain radio-loud and radio-quiet objects, with a division
being made at a ratio of the 5\,GHz radio to the optical $R$-band flux
of 10 (see also Sect.~\ref{sec:rlrq}).  Table~\ref{tab:agn} summarizes
the separation into different types of AGN. Type 0 sources are called
blazars and originally defined sources with no line emission. This
class mainly contains BL Lac objects and flat spectrum quasars
(FSRQs). However, with modern and more sensitive observations emission
lines have been detected in several FSRQs and also in a few BL Lac
objects. Narrow line radio galaxies include two radio morphology
types: the lower-luminosity Fanaroff-Riley type I (FR\,I) radio
galaxies with often symmetrical jets that broaden and fade away from
the nucleus, and the FR\,II radio galaxies, which show highly
collimated jets leading to well-defined lobes with hot spots
\citep{1974MNRAS.167P..31F}.

\begin{table}[htbp]
\caption[Properties of AGN.]{Properties of AGN (see text for details).}
\label{tab:agn}
\centering
\begin{tabular}{l|lll}
\hline\hline
Radio & \multicolumn{3}{c}{Emission line properties}\\
loudness & Type 2 & Type 1 & Type 0\\
         & narrow lines & broad lines & ``no'' lines\\
\hline
radio-quiet: & Seyfert 2 & Seyfert 1 &  \\
            & Liners & Quasar &  \\
radio-loud: & NLRG & Quasar & Blazars \\
\hline
\end{tabular}
\end{table}

\begin{figure}
\centering
\includegraphics[angle=0,width=10cm] {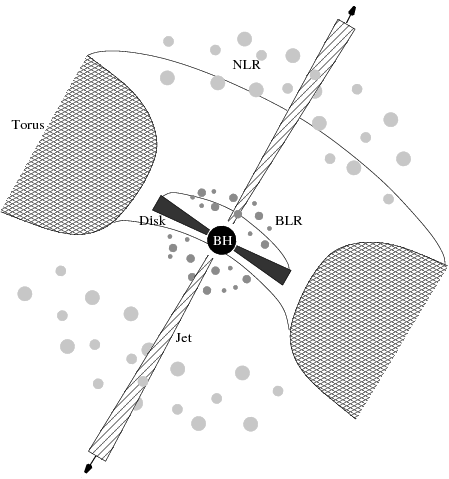}
   \caption[Unification scheme for AGN.]{Unification scheme for AGN (not to scale).
In this picture the appearance of an AGN depends strongly on the viewing angle. For
some directions the central part is hidden by the torus (or an absorber of some
other geometrical form), whereas from other directions the innermost parts are
visible.} 
   \label{fig:unified}
\end{figure}

Obscuring material found in type 2 sources \citep{1977ApJ...213..635R} and the
detection of polarized broad emission lines in Seyfert 2 galaxies
\citep{1985ApJ...297..621A} led to the idea of a unified scheme for AGN. In
addition, depending on the orientation of the jet to our line of
sight, relativistic beaming (see Sect.~\ref{sec:tb} about relativistic
beaming) of the jet emission strongly affects the appearance and
classification of the AGN. Whether AGN are classified type 1 or 2
depends on the obscuration of the luminous nucleus, and whether a
radio-loud AGN is a blazar or a radio galaxy depends on the angle
between the relativistic jet and the line of sight
\citep[e.g.][]{1993ARA&A..31..473A,1995PASP..107..803U}. An
illustration of the unified scheme for AGN is given in
Fig.~\ref{fig:unified}.

\subsubsection{VLBI structures in AGN jets}

AGN are in the focus of VLBI observations since the first measurements
were performed in 1967. At the beginning of VLBI mainly detailed
studies of a few bright sources were carried out, but as new technical
developments increased the observed bandwidth and therefore the
sensitivity of the observations, and with the advent of the VLBA,
larger surveys of hundreds of sources for statistical studies became
possible
\citep{1988ApJ...328..114P,1996ApJS..107...37T,1998AJ....115.1295K,2000ApJS..131...95F,2005AJ....130.1389L}.
Newer surveys even provide total intensity and linear polarization
images of more than 1000 sources
\citep{2002ApJS..141...13B,2007ApJ...658..203H}.

When speaking about relativistic jets in AGN one usually only
considers radio-loud AGN, because they harbour the brightest, largest,
and best studied jets.  Nevertheless, they constitute only about
10\,\% of the AGN population and it seems likely that also radio-quiet
AGN produce jets
\citep[e.g.][]{1996ApJ...473L..13F,1996ApJ...471..106F,2005ApJ...621..123U}. The
formation of powerful jets may depend on many parameters, but given
the number of different types of objects from AGN to stellar black
holes and Herbig-Haro objects, where jets have been observed, jet
formation must be possible in almost all circumstances.

\begin{figure}
\hbox{
\centering
\includegraphics[angle=-90,width=7.2cm] {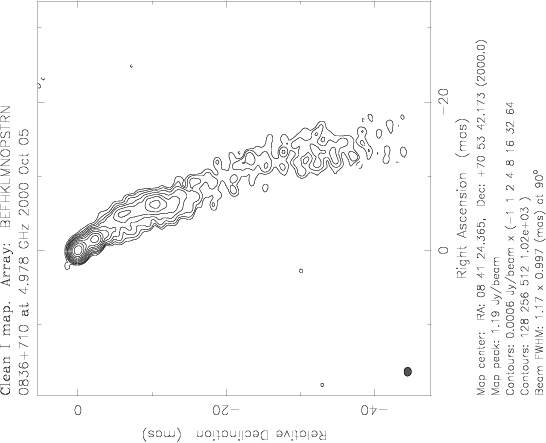}\hspace{2pt}
\includegraphics[angle=-90,width=6.7cm] {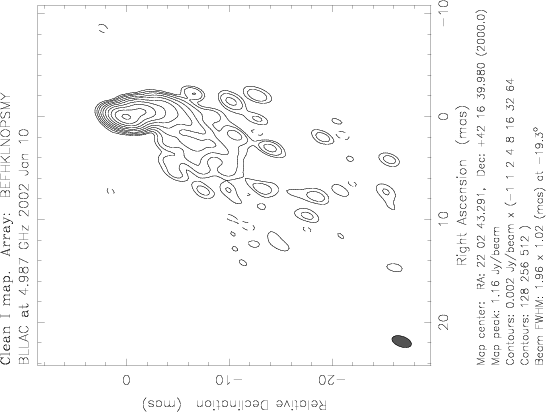}}\vspace{5pt}
\hbox{
\includegraphics[angle=-90,width=5.7cm] {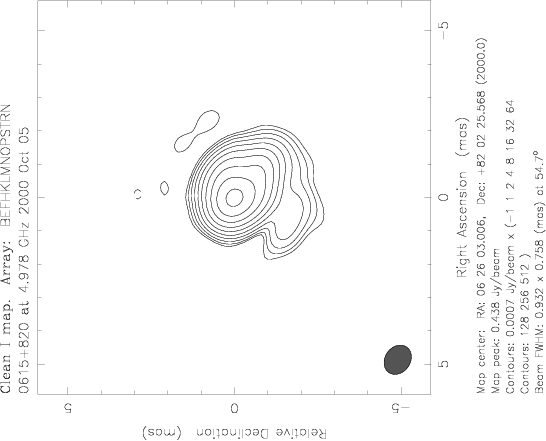}\hspace{2pt}
\includegraphics[angle=-90,width=8.2cm] {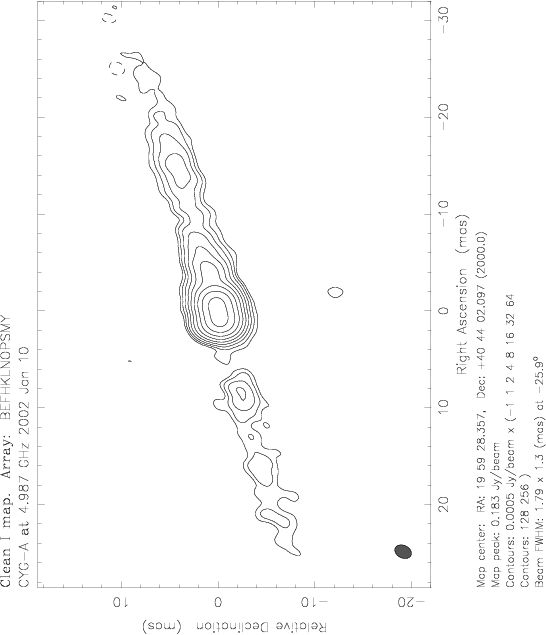}}
   \caption[Typical parsec-scale radio morphology of AGN.]{Typical parsec-scale
radio morphology of AGN seen at 5\,GHz using the VLBA plus Effelsberg. 90\,\% of the
sources show this core-jet structure with smooth parts, knots, bends or gaps (top
panels). Some are quasi point-like (bottom-left), or, if the viewing angle is large
enough, show double sided jets (bottom-right).} 
   \label{fig:0836}
\end{figure}

The typical parsec-scale radio morphology of a radio-loud AGN is a bright
point-like component, referred to as the ``core'' component, and an extension
pointing away from the core: the jet (Fig.~\ref{fig:0836}, top panels). This class
makes up about 90\,\% of the radio-loud AGN. The rest divides into point-like or
double-sided source (Fig.~\ref{fig:0836}, bottom panels) or more irregular sources.
In the case of the point-like sources we probably see only the core, but for the
others it is sometimes more difficult to judge which component the core is. Usually
this is done on the bases of compactness and the radio spectrum. The core has a flat to inverted radio spectrum ($\alpha > -0.5$ with $S\propto
\nu^\alpha$). Since the synchrotron spectrum should be flat only at its peak
frequency the appearance of flat spectra in AGN cores over a wide
range of frequencies caused some discussion
(e.g. \citealt{Marscher1977}). There is still some discussion about
this, but the most common explanation is that the flat core spectrum
is the result of a superposition of several distinct synchrotron
components each of which has a peaked spectrum at a slightly different
frequency (``cosmic conspiracy'', \citealt{Cotton1980}). The jets have
usually steeper spectra with $\alpha<-0.7$, resulting from optically
thin synchrotron emission.

In many aspects the appearance of radio-quiet objects is similar to
their radio-loud counterparts. However, extended jet structures are
rare and besides the core only some distinct jet components are visible
\citep[e.g.][]{2001RvMA...14...15F,2002A&A...392...53N,2005ApJ...621..123U}. The
appearance of parts of the jet and the observations of multiple bend in the jets of
Seyfert galaxies might be explained be the interaction of the jet with NLR or BLR
clouds \citep[e.g.][]{2003ApJ...583..192M,2007MNRAS.377..731M}.

A comparison between the VLBI images in Fig.~\ref{fig:0836} and the
sketch of an AGN in Fig.~\ref{fig:unified} is not always
straightforward. The jet in Fig.~\ref{fig:0836} (top panel) might be
easy to identify, however, the ``core'' for example does not
correspond to the central engine, but usually marks the transition
region of optically thick to thin emission
\citep{1981ApJ...243..700K}.  Since the location of this transition
zone depends on the observing frequency, the apparent location of the
core moves towards the central engine at higher frequencies, which has
been observed in some sources
\citep{1998A&A...330...79L,2005ASPC..340...25M}.  Other possibilities
are that the core is a conical ``re-collimation'' shock that
accelerates particles
\citep{2005MNRAS.357..918B,1995ApJ...449L..19G,1988ApJ...334..539D} or the jet
initially is not well aligned with the line of sight, but changes its
direction towards the observer so that the beaming factor increases
and the jet becomes visible. The formation of the jet itself takes
place presumably on scales of a few 10 to 1000 Schwarzschild radii
($R_{\rm S}= 2GM/c^2$). The Schwarzschild radius of a typical AGN
black hole of $10^8$\,$M_\odot$ is of the order of 2\,AU
($3\times10^{12}$\,m, $R_{\rm S}\approx3000\,{\rm m}
\frac{M}{M_\odot}$).

The closest look so far into the central engine of an AGN was obtained
with global VLBI observations of M\,87 at 43\,GHz and 86\,GHz
\citep{1999Natur.401..891J,2006JPhCS..54..328K,Ly2007}. M\,87
is an E0 galaxy at the centre of the Virgo cluster, and has one of the
nearest extragalactic jets. Together with its large inferred black
hole mass of $3\times10^9$\,M$_\odot$ \citep[e.g.][and references
therein]{1997ApJ...489..579M}, M87 is an ideal candidate for jet
formation studies.  At a distance of 14.7\,Mpc, 1\,mas corresponds to
a linear scale of 0.071\,pc and the Schwarzschild radius of the black
hole is about 60\,AU, corresponding to 0.0003\,pc. Surprisingly, this
is very similar to the linear resolution in Sgr\,A$^*$\footnote{M\,87
is approximately 2000 times more distant than Sgr\,A$^*$ but has a
black hole with a mass 1000 greater than Sgr\,A$^*$'s, so these two
effects almost exactly cancel each other when the resolution is
measured in Schwarzschild radii.}. The observations revealed that the
jet in M87 starts fairly broad at an opening angle of 60$^\circ$ on
scales $<0.5$\,mas (0.04\,pc), but collimates to about 5$^\circ$
within the first few parsecs (e.g.,
\citealt{Ly2007,Kovalev2007}). Together with the slow proper motion in
the parsec-scale jet
\citep{1989ApJ...336..112R,1995ApJ...447..582B}, this supports the
idea of a magnetically dominated launching of the jet with a long
collimation and acceleration zone \citep{2004ApJ...605..656V}. In its
initial phase (ultra compact jet) the jet is probably dominated by
electromagnetic processes
\citep{2001Sci...291...84M,2005ApJ...625...72S}, and is visible
as the compact VLBI-core where the jet becomes optically thin for
radio emission
\citep{1998A&A...330...79L,1999ApJ...521..509L}. 

Several other VLBI observations made use of the space-VLBI technique
to achieve highest angular resolution, and to resolve prominent quasar
jets in the transverse direction (e.g. 3C273:
\citealt{2001Sci...294..128L}; 0836+714:
\citealt{2006PASJ...58..253L}). This allowed a detailed study of jet
propagation and jet internal processes, like i.e. the development of
instabilities. In Centaurus\,A VSOP observations at 5\,GHz by
\citet{2006PASJ...58..211H} revealed a sub-parsec scale jet-opening
and collimation region similar to that observed in M\,87 at mm
wavelengths
\citep{1999Natur.401..891J,2006JPhCS..54..328K}. Although the
resolution was not as high as in the case of M\,87 the observations
suggested that the jet collimation takes place on scales of a few
10\,$R_{\rm S}$ to 1000\,$R_{\rm S}$, which supports MHD disk outflow
models \citep[e.g.][]{2001Sci...291...84M}.

Future space VLBI observations at higher frequencies will provide even
higher angular resolution and will probably be able to image the
collimation region within $100\,R_{\rm S}$ of the central engine in
many more AGN. At 43\,GHz, the angular resolution, as it is planed for
VSOP\,2, will be a few ten micro-arcseconds, closely matching the
possible angular resolution of ground based mm-VLBI at 230 GHz. Images
obtained at different frequencies and with matched angular resolution,
can help to detect spectral and polarization properties of compact
regions, which are not observable otherwise
\citep[e.g.][]{2001MNRAS.320L..49G,2006MNRAS.373..449P,2005MNRAS.356..859P}.

\begin{figure}
\centering
\includegraphics[width=12cm]{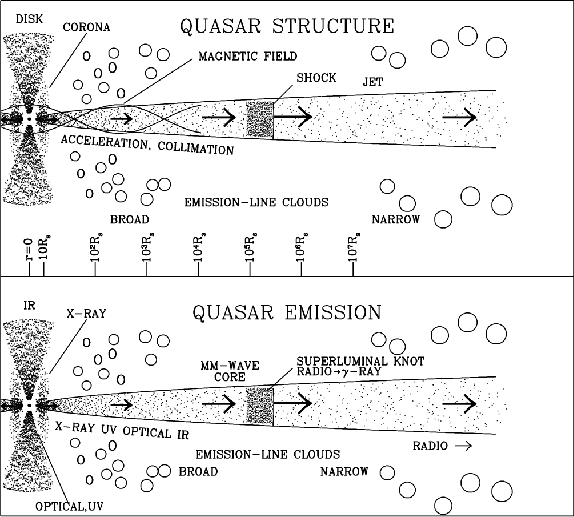}
  \caption[Sketch of a radio-loud AGN.]{Illustration of the central part of a
radio-loud AGN. The density of the dots in the disk, corona, and jet
very roughly indicates the density of plasma (top panel) or intensity of emission
(bottom panel) in a reference frame in which there is no beaming. Note the
logarithmic length scale beyond 10\,$R_{\rm s}$, where $R_{\rm s}$ is the
Schwarzschild radius. Only a single superluminal knot is shown; usually there 
are several \citep[courtesy][]{2005MmSAI..76..168M}.} 
   \label{fig:jetsketch}
\end{figure}

An illustration of the structures and emitting regions of the central
part of a radio-loud AGN is given in
Fig.~\ref{fig:jetsketch}. Observations suggest that the ultra compact
jet is dominated by smooth changes in the particle density associated
with nuclear flares in the central engine and not with strong shocks
\citep{1998A&AS..132..261L,1999ApJ...521..509L}. On the other hand parsec-scales shock
models
\citep{1985ApJ...298..301H,1985ApJ...298..114M,1989ApJ...341...54H}
are able to explain the observed flux density and polarization
characteristics of jets
\citep{2005AJ....130.1418J,2005AJ....130.1389L,2006MNRAS.367..851C}. Further
evidence for shocks comes from observations of rapid changes of the
turnover frequency (frequency at the maximum of the synchrotron
spectrum) along the jet
\citep{1998A&AS..132..261L}.

In several AGN small changes in the jet direction on sub-parsec scales
over a period of a few years have been observed
\citep[e.g.][]{1986ApJ...308...93B,1993A&A...279...83C,2001ApJ...556..738J,2003MNRAS.341..405S,2005AJ....130.1418J}. For
example, global 3\,mm VLBI observations of NRAO\,150 revealed an
apparent rotation of the inner jet at a rate of up to $\sim11^\circ$
per year, which is associated with a non-ballistic superluminal motion
of the jet within this region and probably would have remained
undetected at lower resolution \citep{2007arXiv0710.5435A}.

The most common explanation for this phenomenon is a precessing jet,
which could be caused by a binary super-massive black hole system
\citep{1986ApJ...308...93B,1992A&A...257..489H}.  Although the
observed angles are small ($<10^\circ$), which translates to less than
$1^\circ$ when deprojected, the time scale is set by the movement of
some section of the jet that might not itself move with relativistic
velocities. Therefore, the time scale is unaffected by Doppler
blueshifting and can lead to rather extreme physical conditions of the
nuclear region \citep{Lobanov2005}. The time scale for the precession
of a supermassive black hole is of the order of $\geq 10^4$ years and
becomes visible on kiloparsec scales
\citep{1982ApJ...262..478G}. However on parsec scales hydrodynamic
effects also can lead to a precession of the jet
\citep[e.g.][]{2002ApJ...572..713H,2003ApJ...597..798H}. In this case the
precession may be introduced by a strong shock at the base of the
jet. Observations of more erratic changes in the jet direction
\citep{2005AJ....130.1418J} support this idea and indicate that it is
not necessarily a regular periodic process.

Further out, on scales of about $>100$\,pc, the appearance of the jet
becomes more and more dominated by instabilities of the jet flow, most
importantly Kelvin-Helmholtz instabilities
\citep{2003NewAR..47..629L,2001Sci...294..128L,2006A&A...456..493P,2007A&A...469L..23P}.
Since the jets have steep spectra, high resolution at low frequencies
is necessary to successfully model the transverse structures using
Kelvin-Helmholtz instabilities. Therefore most of those observations
were made at a wavelength of 18\,cm with the Space VLBI Program, which
combines a world array of radio telescopes with the satellite radio
antenna HALCA.

A common method to study the emission processes in the inner jet is
the combined analysis of light curves from radio to X-rays or even
$\gamma$-rays with multi-epoch VLBI observations at high frequencies
\citep[e.g.][]{2001ApJ...556..738J,2002A&A...394..851S,2006A&A...456..105B,2006MNRAS.373.1470P,2007AJ....134..799J}.
The comparison of the broad-band variability with structural changes
in the radio jet allows to directly identify those regions which are
responsible for the variability. The detection of wavelength-dependent
time delays further helps to constrain the distances between the
various emission regions along the jet. In addition, multi-epoch VLBI
observations can be used to measure the jet speed and the source
geometry (see Sect.~\ref{sec:tb}), and the analysis of linear
polarization provides information on the magnetic field structure (see
Sect.~\ref{sec:pol} for more details on VLBI polarimetry).

In more recent observations of 3C\,120 the high resolution of global
3\,mm VLBI was used in combination with its millimetre-wave continuum
spectrum and a computer model to derive the number density of the
combined electron-positron population in the core and compact knots in
the jet \citep{2007ApJ...665..232M}. Assuming that the jet contains a
pure pair plasma, this density and the efficiency of the eventual
annihilation can be used to predict the intensity of the narrow
electron-positron annihilation line, which should be detectable with
modern high energy satellites.

\subsubsection{Brightness temperatures and jet speeds}\label{sec:tb}

The brightness temperature, $T_{\rm B}$, the physical temperature a
blackbody would need to have to produce the observed radio flux
density, is defined in analogy with the Rayleigh-Jeans Law as:

\begin{equation}\label{eq:tbo}
T_{\rm B}=\frac{c^2 B_\nu}{2\nu^2k},
\end{equation}

where $T_{\rm B}$ is given in K, $c$ is the speed of light in m\,s$^{-1}$,  $B_\nu$
is the flux density per unit solid angle, $\Omega$, in
W\,m$^{-2}$\,Hz$^{-1}$\,sr$^{-1}$, $\nu$ is the observing frequency and  $k$ the
Boltzmann constant in J/K. To convert Eq.~\ref{eq:tbo} into convenient units,
$\Omega$ can be expressed as the area of a sphere intersected by a cone with
opening angle $\theta$, yielding $\Omega=4\pi\sin^2(\theta/4)$. Using the
approximation $\sin(x)\approx x$ for $x <<1$ and introducing units of
milli-arcseconds, $\Omega=1.846\times10^{-17}\,\theta$. Converting $B_\nu$ and 
$\nu$ to Jy/beam and GHz, respectively, yields

\begin{equation}
T_{\rm B}=1.76\times10^{12} \frac{S_\nu}{\theta^2\nu^2}.
\end{equation}

VLBI core components typically have measured brightness temperatures
of $10^{10}$\,K to $10^{12}$\,K, which is well above the limit for
thermal emission, and indicates a non-thermal origin such as
synchrotron emission from relativistic electrons. However, even the
brightness temperature for synchrotron radiation has an upper-limit --
the Compton limit. \citet{1969ApJ...155L..71K} have shown that for
brightness temperatures exceeding $10^{12}$\,K the energy losses due
to inverse Compton scattering become dominant and the brightness
temperature decreases to values between $10^{11}$\,K to $10^{12}$\,K,
where inverse Compton and synchrotron losses are of the same order.

Indeed most of early VLBI experiments showed that the brightness
temperatures of nearly all sources falls into the small range of
$T_{\rm B}\approx10^{11}$\,K to $T_{\rm B}\approx10^{12}$\,K
\citep{1969ApJ...155L..71K}. However, in the following decades it
became clear that this was a natural consequence of the size of the
Earth (longest possible baseline) and the limited range of flux
densities observed, and had nothing to do with radio source
physics. This is because the resolution of any interferometer is given
by $\theta=\lambda/D$ where $\lambda$ is the wavelength of
observation, and $D$ is the baseline length. For a typical radio
source with $1\,{\rm Jy} < S_\nu < 10$\,Jy, and a common baseline
length $D \sim5000$\,km to 8000\,km, $T_{\rm B}$ will always be
$10^{11}$\,K to $10^{12}$\,K
\citep{2003ASPC..300..185K}. 

More recent VLBA \citep{1998AJ....115.1295K,2002AJ....124..662Z} and
space VLBI observations
\citep[e.g.][]{2001ApJ...549L..55T,2004ApJ...616..110H} suggest that
there are sources in which the maximum observed brightness temperature
reaches $10^{13}$\,K or even higher values. In particular IDV sources were found to
have considerably higher core brightness temperatures than non-IDV
sources \citep{2001ApJ...549L..55T,2001ApJ...554..964L}, which agrees
with the finding that the IDV originates from the VLBI-core component
\citep{2006A&A...452...83B}. There is evidence that even the majority of the radio-loud AGN have core
brightness temperatures above $10^{12}$\,K
\citep{2004ApJ...616..110H}. Conversely, a lower limit of $T_{\rm B}$
of around $10^{5-6}$\,K is given by the limited sensitivity of today's
VLBI observations of a few 10\,$\mu$Jy to 100\,$\mu$Jy.

\begin{figure}
\centering
\includegraphics[angle=-90,width=12cm] {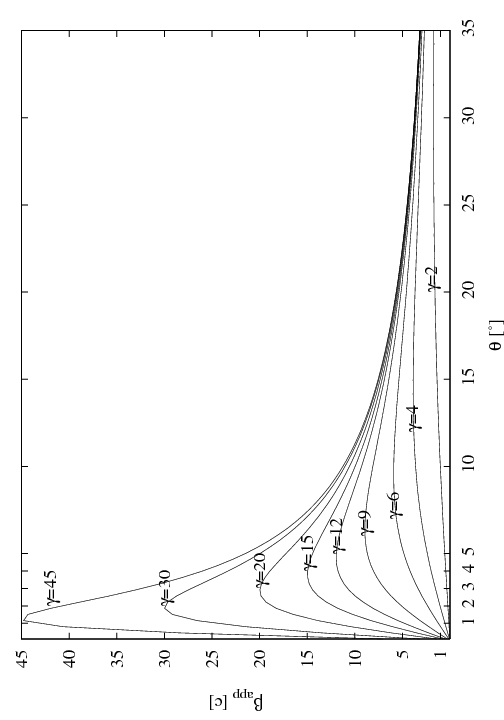}
  \caption[Relation between intrinsic speed, viewing angle, and apparent jet
speed.]{Illustration of the dependence of the apparent jet speed from the viewing
angle and the intrinsic jet speed.} 
   \label{fig:jetspeed}
\end{figure}

The excess of $T_{\rm B}$ is usually explained in terms of
relativistic beaming. If the jet is moving relativistically,
relativistic beaming will increase the flux density without changing
the size of the jet, hence it will lead to an apparent increase of
$T_{\rm B}$ beyond the allowed value, proportional to the Doppler
factor, $\delta$: $T'_{\rm B}\propto T_{\rm B}\,\delta$ with
$\delta=[\gamma(1 - \beta
\cos\theta)]^{-1}$, where $\gamma$ is Lorentz factor of the flow, $\theta$ is the
viewing angle, and $\beta=v/c$ is the speed in units of the speed of
light. This concept had already been discovered by
\citet{1967MNRAS.135..345R} before the first observations of strong
variability
\citep{1965Sci...148.1458D,1965SvA.....9..516S,1966AJ.....71R.394P} and
superluminal motion in radio jets \citep{1971Sci...173..225W,1977Natur.268..405C}
raised the questions for relativistic beaming. Finally the introduction of the twin
relativistic jet model by \citet{1974MNRAS.169..395B} formulated the observational
consequences as:

\begin{equation}\label{eq:beta_app}
\beta_{\rm app}=\frac{\beta \sin\theta}{1-\beta \cos\theta},
\end{equation}

which allows the observer to register superluminal motion in jets if the intrinsic
jet speed, $\beta$, is close to the speed of light and the viewing angle, $\theta$,
is small (see Fig.~\ref{fig:jetspeed}). The maximum speed, $\beta_{\rm
app,max}=\sqrt{\gamma^2 -1}$, occurs when $\cos \theta = \beta$ or $\sin \theta  =
\gamma^{-1}$. At this angle the Doppler factor is equal to the Lorentz factor. For
the flux density one obtains:

\begin{equation}\label{eq:beaming}
S=S_{\rm 0}\delta^{(3-\alpha)},
\end{equation}

where the spectral index $\alpha$ in the exponent corrects for the
Doppler effect, which causes a frequency shift in the spectrum by the
Doppler factor, $\delta$. For a continuous jet the exponent changes to
($2-\alpha$) due to geometrical considerations
\citep{1979Natur.277..182S}.

Following equipartition arguments, \citet{1994ApJ...426...51R} found an even lower
limit to the brightness temperature. The author states that if there is no large
departure between the magnetic and the particle energy (equipartition), then the
limiting brightness temperature is $5\times 10^{10}$\,K to $10^{11}$\,K.

\begin{figure}
\centering
\includegraphics[angle=0,width=6cm] {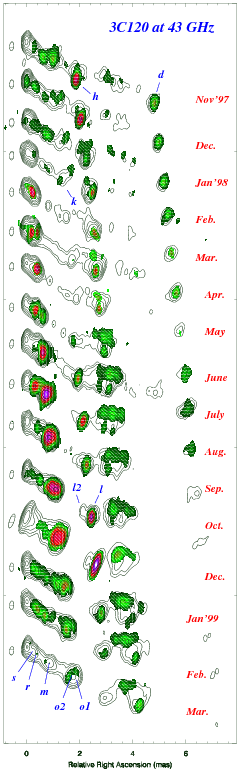}
  \caption[Time series of VLBA images of the jet of the radio galaxy 3C\,120]{Time
series of VLBA images of the jet of the radio galaxy 3C\,120 showing many details
of the structural evolution of the jet. Contours give the
total intensity, colours (on a linear scale from green to white) show the polarized
intensity, and bars (of unit length) indicate the direction of the magnetic
polarization vector. Synthesized beams are plotted to the left of each image, with
a typical size of $0.35 \times 0.16$\,mas \citep[courtesy][]{2001ApJ...561L.161G}.} 
   \label{fig:3c120a}
\end{figure}

\begin{figure}
\centering
\includegraphics[angle=0,width=9cm] {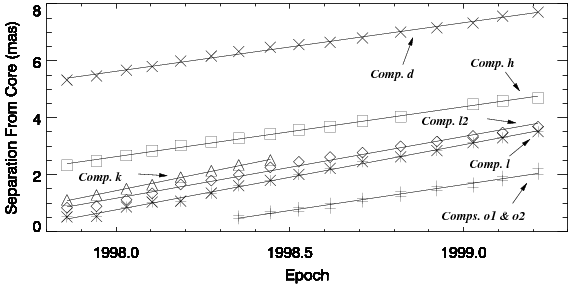}
  \caption[Projected angular distance from the core as a function of time for the
jet of 3C120.]{Projected angular distance from the core as a function of time for
the superluminal components in the jet of 3C120. Those are the components which
could be traced over the whole period shown in Fig.~\ref{fig:3c120a} \citep[courtesy][]{2001ApJ...561L.161G}.} 
   \label{fig:3c120b}
\end{figure}

Indeed jet speeds of up to 40 times the speed of light, corresponding to a minimum
Lorentz factor of 40, have been measured
\citep[e.g.][]{2001ApJS..134..181J,2006ApJ...640..196P}. Studies of the motion of
AGN jet components is one of the main research fields of VLBI. In
Figures~\ref{fig:3c120a} and \ref{fig:3c120b} a time series of VLBA
images taken at 7\,mm and the corresponding identification of
superluminal jet components in the jet of the radio galaxy 3C\,120 are
shown. Recently, detailed studies of individual sources are
accompanied by large, statistically relevant surveys
\citep{2001ApJ...556..738J,2004ApJ...609..539K,2005ASPC..340...20L}. So far, VLBI
at radio frequencies is the only tool to spatially resolve the jets on
parsec scales. Compact interferometers such as the VLA and MERLIN
provide information on kilo-parsec scale jet motion
\citep[e.g.][]{2002MNRAS.336..328L} and in the optical the {\it Hubble} Space Telescope was
used to measure motions in the optical jet of M\,87
\citep{1999ApJ...520..621B}. Although the most famous and usually also the
brightest sources are those showing high superluminal motion with $\beta_{\rm
app}>10\,c$, most of the radio jets detected display much lower speeds of less then
$3\,c$ \citep{2004ApJ...609..539K,2005ASPC..340...20L,2007ApJ...658..232C}. 

This is commonly interpreted as a result of the relativistic beaming
model and reflects the fact that the sources are orientated at
different angles to the line of sight.  However, many of the
low-velocity sources are high-power objects like quasars and BL\,Lac
objects. Their rapid variability and high brightness temperatures
suggest that the bulk motion of the jet is highly relativistic. Some
of them might have very small viewing angles corresponding to the
region left of the peaks in Fig.~\ref{fig:jetspeed}, which reduces the
apparent speed, but the probability that they are all oriented at very
small angles to the line of sight is quite small
\citep[e.g.][]{2007ApJ...658..232C}. It is more likely that the observed speed in
those jets does not correspond to the maximum speed of the flow itself. Indeed
there are a number of sources where components at different speeds are observed,
something which cannot be explained by jet bending or variable Lorentz factors. In several
other cases the speeds measured at higher spatial resolution are substantially
higher than the measured speeds at lower resolution
\citep[e.g.][]{2001ApJ...556..738J,2004ApJ...609..539K}.

For radio galaxies which generally show lower jet speeds increased
evidence for structured jets is found. In this case the jet is
expected to consist of a fast spine whose radio radiation is beamed
away from our line of sight, surrounded by a slower sheath
\citep[e.g.][]{1999MNRAS.306..513L,2001ApJ...549L.183A,2002MNRAS.336..328L,2004ApJ...609..539K,2007ApJ...658..232C}.
Seyfert galaxies, which usually display motions of $<0.5\,c$, are
supposed to have initially weaker jets, and they often show signs of
interaction with the surrounding medium \cite[e.g.][and references
therein]{2003ASPC..300...97U}.  Nevertheless, superluminal motion has
been observed in some sources after a bright outburst
\cite[e.g.][]{2000A&A...357L..45B}.

\subsubsection{The radio-loud / radio-quiet divide}\label{sec:rlrq}

Radio-loud AGN are sources where the ratio of the 5\,GHz radio flux
density to the optical $R$-band flux density is of the order of 10 or
larger. Sources with lower ratios are called radio quiet. Only about
10\,\% of the AGN population are radio-loud objects, nevertheless, most
of the VLBI studies are dealing with those sources. VLA and VLBI
observations of radio-quiet sources revealed that their appearance is
quite similar to radio-loud objects exhibiting cores, jets, and
lobe-like structures, but just a bit weaker
\citep[e.g.][]{1994AJ....108.1163K,2005ApJ...621..123U}. Also the spectral
characteristics are similar, and variability studies comparing samples
of radio-loud and radio-quite objects revealed that the variability
time scales are independent of radio luminosity or radio-to-optical
flux density ratio. This all leads to the conclusion that the physics
of radio emission in the inner regions of all quasars is essentially
the same, involving a compact, partially opaque core together with a
beamed jet \citep[e.g.][]{2005ApJ...618..108B}. The formation of
powerful jets might depend on a fine tuning of several parameters like
the spin of the black hole, the type of accretion flow, and a
favourable magnetic field geometry
\citep[e.g.][]{2007MPLA...22.2397B}. Currently many authors seem to
prefer the spin theory. \citet{1995ApJ...438...62W} suggested that the
coalescence of two super-massive black holes (SMBHs) after the merger
of two massive galaxies might spin up the black hole so that it can
produce a powerful jet. If not the coalescence of SMBHs, at least the
merger history of a galaxy seems to play an important role for the
radio-loudness \citep{2006A&A...453...27C}. Although they are harder
to image radio-quiet quasars provide important information on the
launching of jets and are crucial for the development of a unified
scheme for active galaxies.

\subsubsection{Intraday variability}

Since the discovery of intraday variability (IDV, i.e. flux density and
polarization variations on time scales of less than 2 days) about 20 years ago
\citep{1986MitAG..65..239W,1987AJ.....94.1493H}, it has been shown that IDV is a
common phenomenon among extra-galactic compact flat-spectrum radio sources. It is
detected in a large fraction of this class of objects
\citep[e.g.][]{1992A&A...258..279Q,2001MNRAS.325.1411K,2003AJ....126.1699L}. The
occurrence of IDV appears to be correlated with the compactness of the
VLBI source structure on milli-arcsecond scales: IDV is more common and
more pronounced in objects dominated by a compact VLBI core than in
sources that show a prominent VLBI jet
(\citeauthor{1992A&A...258..279Q} 1992; see also
\citeauthor{2001ApJ...554..964L} 2001). In parallel to the variability of the total
flux density, variations in the linearly polarized flux density and the
polarization angle have been observed in many sources 
\citep[e.g.][]{1989A&A...226L...1Q,1999bllp.conf...49K,1999NewAR..43..685K,2003A&A...401..161K,2004ChJAA...4...37Q}.
The common explanation for the IDV phenomenon at cm wavelengths is
interstellar scattering
\citep[e.g.][]{1995A&A...293..479R,2001Ap&SS.278....5R}.  The most
convincing argument for interstellar scintillation comes from
observations of time-delays between the IDV pattern from the same
source arriving at two widely separated telescopes
\cite{2002Natur.415...57D}. Another frequently observed phenomenon is the
annual modulation of the variability time scale. This is interpreted
in terms of the change of the relative velocity between the scattering
screen and the velocity of the Earth as it orbits the Sun
\cite[e.g.][]{2001ApJ...550L..11R,2003ApJ...585..653B,2003A&A...404..113D,2007A&A...470...83G}. 
However some effects remain that cannot be easily explained by
interstellar scintillation and that are probably caused by the
relativistic jets.
\citep[e.g.][]{1996ChA&A..20...15Q,2002ChJAA...2..325Q,2004ChJAA...4...37Q}. For
example the correlated intra-day variability between radio and optical wavelengths,
which is observed in sources such as 0716+714 and 0954+658, suggests that at least
part of the observed IDV has a source-intrinsic origin 
\citep[e.g.][]{1990A&A...235L...1W,1991ApJ...372L..71Q,1996AJ....111.2187W}. Also
the recent detection of IDV at millimetre wavelengths in 0716+714 
\citep{2002PASA...19...14K,2003A&A...401..161K,2006A&A...456..117A} is a problem
for the interpretation of IDV by interstellar scintillation, because
scintillation decreases with the square of the observing frequency.

Independent of the physical cause of IDV (source intrinsic, or induced
by propagation effects), it is clear that IDV sources must contain one
or more ultra-compact emission regions. Using scintillation models,
typical source sizes of a few tens of micro-arcseconds have been
derived
\citep[e.g.][]{1995A&A...293..479R,2002Natur.415...57D,2003ApJ...585..653B}. In the
case of source-intrinsic variability and when using arguments of light
travel times, even smaller source sizes of a few micro-arcseconds are
obtained. Then apparent brightness temperatures of up to
$10^{18-19}$\,K (in exceptional cases up to $10^{21}$\,K) are
inferred, far in excess of the inverse Compton limit of
$10^{12}$\,K. These high brightness temperatures can be reduced by
relativistic beaming with high Doppler-factors
\citep[e.g.][]{1991A&A...241...15Q,1996ChA&A..20...15Q,2002PASA...19...77K}. It
is currently unclear if Doppler-factors larger than 50 to 100 are
possible in compact extragalactic radio sources (see
Sect.~\ref{sec:tb} for more details on observed jet speeds). Space
VLBI observations revealed that the variability in 0716+714 originates
from the core, but the core itself remains unresolved and no structural
changes were observed on short time scales
\citep{2006A&A...452...83B}. However, future space VLBI missions or
millimetre VLBI should be able to resolve the inner structures and
shed more light on the very compact regions of IDV sources.

\subsubsection{Polarization in AGN jets}\label{sec:pol}

In the commonly accepted picture where accretion onto a black hole powers the AGN,
it is believed that the jet is driven outwards via magnetic forces. The poloidal
magnetic field is wound up by the accretion disk or the ergosphere of the black
hole and therewith accelerates and collimates the jet
\citep[e.g.][]{2001Sci...291...84M}. Therefore important information about the
physical properties of the jet are provided by the order and the
geometry of the magnetic field. Since the emission process in the
radio band is synchrotron radiation, the most direct way to
investigate magnetic fields is to study the linear polarization of
jets. The first studies of linear polarization using VLBI observations
were already performed more than 20 years ago
\citep{1984ApJ...286..503C}. More recent studies dealing with larger samples of AGN
have dramatically increased our knowledge about the linear polarization properties
of the various types of AGN
\citep[e.g.][]{1993ApJ...416..519C,1996MNRAS.283..759G,2000MNRAS.319.1109G,2000ApJ...541...66L,2002ApJ...577...85M,2003ApJ...589..733P,2005AJ....130.1389L,2007AJ....134..799J}.
The degree of linear polarization of the VLBI core of a typical
radio-loud AGN is usually below 5\,\%, because the core region is
optically thick and the maximum expected degree of linear polarization
for such a region is 10\,\%. In the optically thin regime, where most of
the emitted synchrotron photons can escape from the source without
being absorbed, the maximum degree can reach 75\,\%
\citep{1966MNRAS.133...67B}. Indeed the observed degree of linear polarization in
the jet reaches up to 50\,\%, confirming its origin to be optically thin synchrotron
emission \citep[e.g.][]{2005AJ....130.1389L}. Usually the linear polarization in
the jet is enhanced at the position of bright jet components (see also
Fig.~\ref{fig:3c120a}) suggesting the existence of shocks that increase the order
of the magnetic field \cite[e.g.][]{2000ApJ...541...66L}.

At the transition from optically thick to thin synchrotron emission
the polarization angle is supposed to rotate by 90$^\circ$. This was
observed in a number of sources at lower frequencies \citep[e.g.\
OJ287][]{2001MNRAS.320L..49G}.  The difficulty here and for the
measurement of the electric vector position angle (EVPA) in general is
that the radio polarization angle can be affected by Faraday rotation
\citep{Faraday1933}. This is a rotation of the EVPA which occurs when a
linearly polarised electromagnetic wave travels through a region with
free electrons and a magnetic field with a non-zero component along
the line of sight. The intrinsic polarization angle $\chi_0$ is
related to the observed polarization angle $\chi$ by

\begin{equation}
\chi = \chi_0 + {\rm RM} \lambda^2
\end{equation}

where $\lambda$ is the observed wavelength and RM is the rotation
measure.  The linear relationship to $\lambda^2$ is the characteristic
signature of Faraday rotation. The rotation measure depends linearly
on the electron density, $n_{\rm e}$, the component of the magnetic
field parallel to the line of sight, $B_\parallel$, and the path
length, $l$, through the plasma. Using units of m$^{-3}$, Tesla, and
parsec, the rotation measure can be expressed as

\begin{equation}
{\rm RM} = 8120 \int n_{\rm e} B_\parallel dl
\end{equation}

The RM can be determined using simultaneous multi-frequency
observations, which then allow the recovery of the intrinsic
polarization orientation.

\citet{2003ApJ...589..126Z,2004ApJ...612..749Z} have conducted an RM study of about
40 AGN. The authors found that the cores of quasars have typical RMs
of approximately 500\,rad\,m$^{-2}$ to several 1000\,rad\,m$^{-2}$
within 10\,pc of the core. Jet RMs are typically 500\,rad\,m$^{-2}$ or
less. The cores and the jets of the seven BL\,Lac objects have RMs
comparable to those of quasar jets. Radio galaxies were usually found
to have depolarized cores and exhibit RMs in their jets varying from a
few hundred to 10\,000\,rad\,m$^{-2}$. A gradient in the foreground
Faraday screen is invoked to explain the observed depolarization
properties of the sample. The Faraday screen is likely located close
to the relativistic jet, although its exact nature remains
unclear. The line-of-sight magnetic fields inferred from the
observation range from 10\,pT to 60\,pT in jets and to 100\,pT in
quasar cores.

The cores of quasars show higher RMs than BL\,Lac objects, but,
conversely, show lower degrees of linear polarization
\citep[e.g.][]{2005AJ....130.1389L}. A possible explanation is that at
least part of the missing linear polarization in quasars is removed by
Faraday depolarization, where the propagation of emission through
different patches of a turbulent plasma produces stochastic Faraday
rotation, which leads to an overall reduction of the observed linearly
polarized component. Also the absence of linear polarization in radio
galaxies might be attributed to Faraday depolarization. Since radio
galaxy jets are often observed in the plane of the sky the cores are seen
through the dense environment surrounding the core region
\citep{2004ApJ...612..749Z}.

When corrected for the rotation measure the EVPA can give an estimate for the
orientation of the magnetic field in the jet. It turns out that BL\,Lac objects
tend to have EVPAs parallel to the jet direction, but for quasars the  correlation
is not that obvious
\citep[e.g.][]{1993ApJ...416..519C,2000MNRAS.319.1109G,2003ApJ...589..733P}. For
optically thin emission this means that the magnetic field is orientated transverse
to the jet direction. Initially this was interpreted as evidence for transverse
shocks which enhance the magnetic field component in this direction
\citep[e.g.][]{1989ApJ...341...54H}. However, the effect can also be explained
in terms of helical magnetic field structures, which are likely to be
present in the inner jet \citep{2005MNRAS.360..869L}. Larger
statistics and the correlation with properties from other wavelengths
appear to be the methods that will help to disentangle the different
effects \citep[e.g.][]{2005AJ....130.1389L,2007AJ....134..799J}.

Another option to probe the particle population and magnetic field structure of the
jets are studies of circular polarization (CP) on parsec scales. Early single dish
observations showed that only marginal circular polarization ($<0.1\,\%$) is present
in AGN jets \citep{1983ApJS...52..293W,1984MNRAS.208..409K}. Interferometric
studies, however, detected CP in the cores of a number of sources
\citep{1998Natur.395..457W,1999AJ....118.1942H,2001ApJ...556..113H,2003Ap&SS.288...29H,2004ApJ...602L..13H}.
A recent VLBI survey revealed circular polarization in the cores of 20
of a total of 133 observed AGN \citep{2006AJ....131.1262H}. In two
sources CP emission was detected also from separated jet
components. Compared to the integrated values measured from single
dish observations, VLBI images show higher levels of CP. About 15\,\%
of the cores have more than 0.3\,\% of circular polarization and
0.3\,\% to 0.5\,\% were found in jet components. Circular polarization
can be produced in two ways: as an intrinsic component of the emitted
synchrotron radiation or via Faraday conversion of linear to circular
polarization \citep{1977A&A....61..291J}. The current statistics are
not sufficient to determine the origin of CP, but when the intrinsic
mechanism dominates, jets must be a predominantly electron-proton
plasma and must contain a significant fraction of unidirectional
ordered magnetic field. If the conversion mechanism dominates,
circular polarization determines the low-energy end of the
relativistic particle distribution
\citep[e.g.][]{2003Ap&SS.288..143W}, a key parameter in studying the
bulk kinetic luminosity of AGN jets and their particle content
\citep{1993MNRAS.264..228C,1998Natur.395..457W}. Since the composition of jets is
one of the fundamental question in AGN research a lot of effort is
currently put into the accurate determination of circular polarization
in jets.

\subsubsection{VLBI observations of circumnuclear tori}

The previous sections have shown that AGN whose radio axes are close
to the plane of the sky are the more difficult objects to study,
because their jets are weaker, i.e. are less beamed, have low degrees
of polarization and the central engine is highly obscured, which
decreases the amount of additional information from optical and
infrared wavelengths. However, in the radio, the obscuring material
can be used to study the circumnuclear structures. There are mainly
three observational methods: (i) molecular gas seen either masing (so
far only water masers at 1.3\,cm have been detected in AGN) or in
absorption, (ii) atomic gas seen in absorption, and (iii) ionised gas
revealed through free-free absorption.

\begin{enumerate}

\item Water masers are found predominantly in Seyfert 2 or LINER
galaxies and are currently the only resolvable tracers of warm dense
molecular gas in the inner parsec of AGN
\citep[e.g.][]{1996ApJS..106...51B,2004ApJ...617L..29B,2002A&A...383...65H,2003ApJ...582L..11G,2005A&A...436...75H,2006ApJ...638..100K,2006A&A...450..933Z}
Because they are associated with nuclear activity, the most likely
model for exciting the maser emission is X-ray irradiation of
molecular gas by the central engine
\citep[e.g.][]{1994ApJ...436L.127N}. In Mrk\,348 the masers are
located close to the receding jet, suggesting that the masers are
likely to arise from the shock in front of the jet, rather than from
continuum emission \citep{2003ApJ...590..149P}. About 60 AGN
harbouring water masers are known up to date, but not all of them are
bright enough to be observed with VLBI. The most relevant are
NGC\,4258
\citep{1995Natur.373..127M}, NGC\,1386 \citep{1997BAAS...29R1374B}, NGC\,4945
\citep{1997ApJ...481L..23G}, NGC\,1068 \citep{1997Ap&SS.248..261G}, NGC\,3079
\citep{2005ApJ...618..618K,2007MNRAS.377..731M}, IC\,2560
\citep{2001PASJ...53..215I}, and Circinus
\citep{2003ApJ...590..162G}. In these objects the maser emission appears
to trace a nearly edge-on disk of molecular gas at distances of
0.1\,pc to 1\,pc from the supermassive black hole, which is supposed
to be directly connected to the accretion disk.

\begin{figure}
\centering
\includegraphics[angle=0,width=12cm] {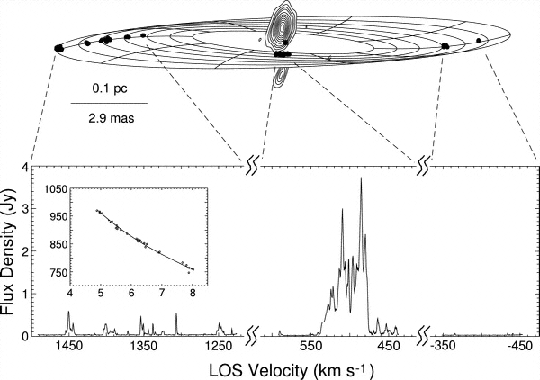}
  \caption[Warped-disk model of NGC\,4258.]{{\bf Top:} Warped-disk
  model for the accretion disc in NGC\,4258 with masers and continuum
  emission superposed. {\bf Bottom:} Total power spectrum of the NGC
  4258 maser, with best-fitting Keplerian rotation curve in the
  inset. Reprinted by permission from Nature
  \citep[][]{1999Natur.400..539H}, copyright (1999) Macmillan
  Publishers Ltd.}  \label{fig:ngc4258}
\end{figure}

The modelling of such maser sources can be used to determine
parsec-scale accretion disk structures and to accurately estimate
black hole masses
\citep{1997Ap&SS.248..261G,2003ApJ...590..162G,2005ApJ...629..719H}. 
Figure~\ref{fig:ngc4258} shows the maser found in NGC\,4258, which can
be well fitted by a Keplerian rotation curve. When combined with the
masers' proper motion or drifts in the line-of-sight velocities of
spectral features (i.e., centripetal acceleration), those models can
provide independent distance measurements to the galaxies and
therewith even provide constraints for $H_0$
\citep{1999Natur.400..539H}. A recent VLBI study aims to measure the
distance to the LINER NGC\,4258 with an error of only 3\,\% and, since
the galaxy also contains Cepheid variable stars, will allow a direct
comparison of the maser and Cepheid distances
\citep{2007ApJ...659.1040A}. \\

\item Evidence of a circumnuclear torus of atomic gas has been seen in Cygnus\,A
\citep{1999ASPC..156..259C}, NGC\,4151 \citep{1995MNRAS.272..355M} and 1946+708
\citep{1999ApJ...521..103P} and many young compact sources
\citep[e.g.][]{1996ApJ...460..634R,2003A&A...404..861V}. In Cygnus A, H\,I
absorption measurements with the VLBA indicate a torus with a radius
of about 50\,pc (Fig.~\ref{fig:cyga}). In NGC\,4151, H\,I absorption
measurements using MERLIN indicate a torus $\sim 70$\,pc in radius and
$\sim 50$\,pc in height. One of the best examples of this type of
torus is the Compact Symmetric Object (CSO) 1946+708. The H\,I
absorption is found in front of all of the $\sim 100$\,pc of its
continuum emission, and the line widths, velocities and H\,I optical
depth distribution are consistent with the scenario of a thick torus,
consisting predominantly of atomic gas
\citep{1999ApJ...521..103P,2001ApJ...554L.147P}.

\begin{figure}
\centering
\includegraphics[angle=0,width=12cm] {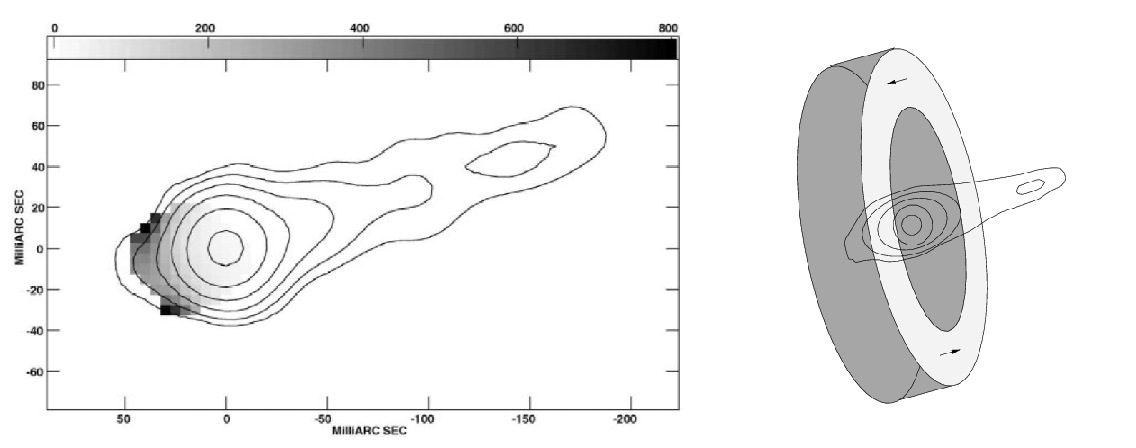}
  \caption[H\,I absorption in Cygnus\,A.]{{\bf Left:} Total intensity contours of
the jet of Cygnus\,A with the H\,I opacity in greyscale
\citep[courtesy][]{1999NewAR..43..509C}. {\bf
Right:} The H\,I absorption is interpreted as the signature of a surrounding torus
\citep[courtesy][]{1999ASPC..156..259C}.} 
   \label{fig:cyga}
\end{figure}

Because radio galaxies often have weak cores and jets, studies of circumnuclear
tori can only be carried out for sources with compact jets. One of the less powerful
cases is the galaxy NGC\,4261, where VLBI observations of H\,I absorption revealed
the existence of a thin disk seen in projection against the counter-jet
\citep{2000A&A...354L..45V}. This shows that there are differences in the nuclear
torus/disk system, which are important parameters for our
understanding of AGN.  Therefore this is one of the science drivers
for more sensitive instruments like the SKA.

\item Parsec-scale radio counter-jets are important for studying the intrinsic
symmetry of the jet-formation process, and are probes of the structure of ionised
gas in the central parsec of AGN. The latter case becomes noticeable as free-free
absorption in front of the counter-jet. If geometrically thin, the disk will cover
only the inner part of the counter-jet but not the approaching jet. If geometrically
thick, the core and perhaps the base of the approaching jet may also be absorbed.
Multi-frequency observations can detect the highly inverted spectrum created by
free-free absorption. This has been found in a number of cases. For example in
3C\,84 \citep{2000ApJ...530..233W} and in the aforementioned case of NGC\,4261
\citep{2001ApJ...553..968J} where H\,I absorption as well as free-free absorption
has been found.

In the FR\,II radio galaxy Cygnus\,A multi-frequency VLBI observations
revealed a deep gap between the core and the counter-jet at lower
frequencies, which changes to an inverted spectrum at high frequencies
\citep{1998A&A...329..873K,2005ASPC..340...30B}. Since  H\,I absorption
was found at the base of the counter-jet
\citep[Fig.~\ref{fig:cyga},][]{1999ASPC..156..259C} the gap most likely indicates
free-free absorption.

\end{enumerate}

\subsubsection{The Galactic Centre}\label{sec:sgra}

The compact radio source at the centre of the Milky Way was found by
\citet{1974ApJ...194..265B} and was named Sagittarius\,A$^*$
\citep[Sgr\,A$^*$,][]{1982ApJ...262..110B} to distinguish it from the larger-scale
galactic region Sgr\,A. A comprehensive review on the galactic centre
and Sgr\,A$^*$ was recently given by \citet{2001ARA&A..39..309M}. The
association with a supermassive black hole had already been suggested
before its discovery by \citet{1971MNRAS.152..461L}, and was
subsequently supported by a number of observations, e.g, by proper
motion measurements of stars in the vicinity of the galactic centre
\citep{1996Natur.383..415E,1997MNRAS.284..576E,1998ApJ...509..678G}
and from size estimates from VLBI observations
\citep[e.g.][]{1994ApJ...434L..59R,1998A&A...335L.106K,2001AJ....121.2610D,2004Sci...304..704B,2005Natur.438...62S,2006ApJ...648L.127B}.
Because of the proximity of Sgr\,A$^*$ high-resolution VLBI images
provide a very high linear resolution. Adopting a distance of 8\,kpc
and a mass of $4\times10^6$\,M$_\odot$
\citep{2003ApJ...597L.121E,2005ApJ...620..744G}, 0.1\,mas correspond
to 10\,$R_{\rm S}$ (Schwarzschild radii, 1$R_{\rm
S}=1.2\times10^{12}$\,cm). VLBI observations of Sgr\,A$^*$ can
therefore help to distinguish between the many different models for
the emission, accretion, and outflow physics of the source, as well as
provide an important test of strong-field gravity
\citep[e.g.][]{2000ApJ...528L..13F}.

The main problem in determining the size of Sgr\,A$^*$ is that its true structure is
affected by scattering in the interstellar medium, leading to a $\lambda^2$
dependence of its diameter as a function of the observed wavelength. Up to 43\,GHz
this results in an elliptical profile orientated at an angle of $\sim 80^\circ$,
the axes of which follow $\theta_{\rm min}=0.76$\,mas\,$(\lambda/{\rm cm})^2$ and
$\theta_{\rm max}=1.42$\,mas\,$(\lambda/{\rm cm})^2$ \citep{1998ApJ...508L..61L}.
At 43\,GHz and higher frequencies there is growing evidence that the intrinsic
source size becomes visible
\citep[e.g.][]{1994ApJ...434L..59R,1998A&A...335L.106K,2001AJ....121.2610D,2006JPhCS..54..328K}.

More recent efforts using the technique of closure amplitudes
\citep[e.g.][]{2001AJ....121.2610D}, which reduces sensitivity but removes
uncertainties due to calibration errors, have resulted in a more robust detection
of the intrinsic source size of Sgr\,$^*$.
\citet{2004Sci...304..704B,2006ApJ...648L.127B} and \citet{2005Natur.438...62S}
determined diameters of $\sim10\,R_{\rm S}$ to $\sim13\,R_{\rm S}$ at 86\,GHz,
which is consistent with an estimate of $11\pm6\,R_{\rm S}$ from earlier VLBI
measurements at 215\,GHz \citep{1998A&A...335L.106K,2006JPhCS..54..328K}.  The peak
brightness temperature was estimated to be $10^{10}$\,K at 86\,GHz, which mainly
excludes advection-dominated accretion flows \citep{1998ApJ...492..554N} and
Bondi-Hoyle accretion models \citep{1994ApJ...426..577M}. Because of the limited
sensitivity in the minor axis size of the scattering ellipse the current
measurements are not able to distinguish between jet models
\citep{1993A&A...278L...1F}, generic radiatively inefficient accretion flows
\citep{2000ApJ...539..809Q}, and hybrids of these models
\citep{2002A&A...383..854Y}. Future arrays including telescopes such as the Large
Mexican Telescope (LMT), APEX, and ALMA, which provide a better uv-coverage and
increased performance at frequencies above 90\,GHz will put further constrains on
the minor axis and might be able to image the shadow of the black hole itself
\citep[e.g.][]{2000ApJ...528L..13F}.

\subsection{Stars}

Stars are not the primary targets for VLBI observations, because their
brightness temperatures (10$^3$\,K to 10$^4$\,K) are much less than
those required for VLBI detections (10$^6$\,K). Nevertheless,
observations of phenomena associated with stars can be made in a few
cases, which will be reviewed in this section.

\subsubsection{Circumstellar disks and envelopes}

Mira-type variable stars are frequent targets of VLBI
observations. Mira stars have a few solar masses and have reached the
end of their lives where the outer envelopes become very extended. The
envelope becomes cool enough for molecules to form, and hence maser
emission can be observed in these objects.

One of the best-studied Mira stars is TX\,Cam, the SiO $v=1$, $J=1-0$
maser emission of which has been observed by \cite{Diamond2003} over a
period of almost two years, corresponding to a full pulsation
period. When combined into a
movie\footnote{http://www.journals.uchicago.edu/ApJ/journal/issues/ApJ/v599n2/58473/video2.mpg,
or http://www.nrao.edu/pr/1999/txcam/txcam.a-ak.gif}, the 44 images,
showing the ring-like maser emission of TX\,Cam in biweekly intervals,
are a fascinating documentary of the evolution of a star. It was found
that the shell from which the emission arises is mostly non-symmetric,
and that gas on outward trajectories shows evidence for gravitational
deceleration. Further observations by
\cite{Yi2005} indicated that the red- and blueshifted masers were
evenly distributed around the star, indicating that it is not
rotating.

Polarimetric observations of SiO masers in Mira, U\,Ori, and R\,Aqr
reported by \cite{Cotton2006} indicate that masing material is being
dragged by magnetic fields in the star. Furthermore, the authors
report possible rotation for only Mira, though this finding could be
affected by Mira's binary companion.

\begin{figure}
\includegraphics[width=\linewidth]{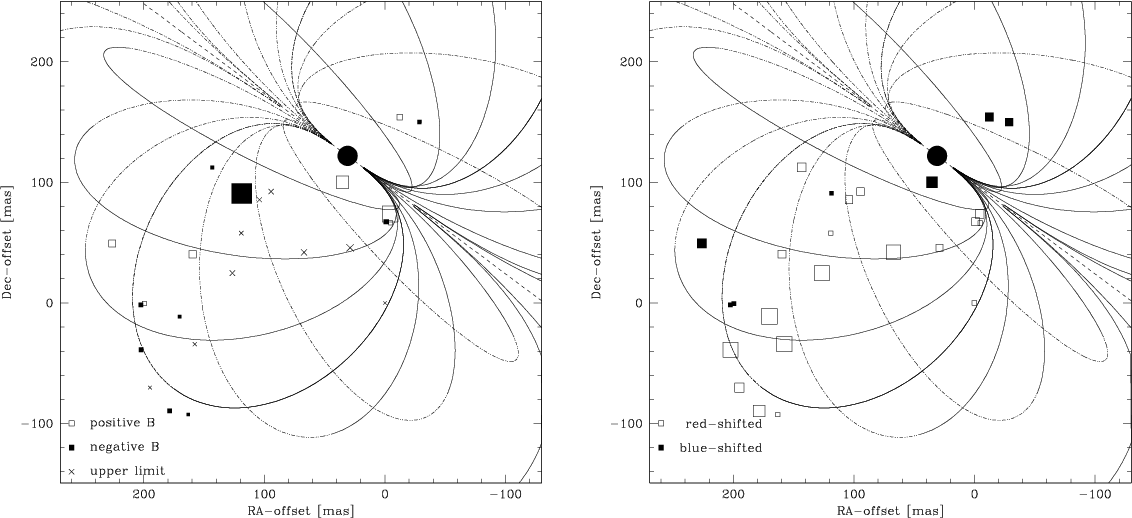}
\caption{A sketch of the dipole magnetic field in VX\,Sgr as derived
from H$_2$O maser observations by \cite{Vlemmings2005}. The left panel
shows the magnetic field orientation (positive or negative) and the
symbol sizes scale with the magnetic field strength. The right panel
shows the velocity distribution of the maser features, showing red-
and blueshifted components. Symbol sizes indicate the velocity
difference. Reproduced with kind permission of the author.}
\label{fig:vlemmings}
\end{figure}

The envelopes of evolved stars can be observed also using H$_2$O
masers. In particular, \cite{Vlemmings2002} and \cite{Vlemmings2005}
report measurements of circular polarization of H$_2$O masers in
evolved stars to determine stellar magnetic fields via the Zeeman
effect. They report magnetic field strengths of several hundred
milligauss to one gauss, and have modelled the stellar magnetic fields
as dipoles (Figure~\ref{fig:vlemmings}).

OH masers trace gas with much lower densities than SiO and H$_2$O
masers. \cite{Claussen1999} and \cite{Hoffman2005} report VLBI
detections of OH masers in, e.g., W28, which \cite{Hoffman2005}
ascribe to the interface between the supernova remnant and a molecular
cloud. From Zeeman splitting they infer magnetic fields of the order
of one milligauss in W28.

At the other end of stellar life cycles, maser observations can yield
information about star formation. Magnetic fields are thought to play
important roles in the formation of massive stars, and have
consequently been observed via polarimetric VLBI observations of
H$_2$O masers, e.g., in Cepheus\,A (\citealt{Vlemmings2006}), W51\,M
(\citealt{Leppanen1998}), and Orion\,KL and W3\,IRS\,5
(\citealt{Imai2003}).

\subsubsection{Stellar jets}

It has been postulated that magnetic fields confine jets of evolved
stars. This could explain the asymmetric shapes of planetary nebulae,
the remnants of stars. \cite{Vlemmings2006a} present evidence for
magnetic confinement of the jet of W43A, using polarimetric VLBI
observations of its H$_2$O maser emission.

\subsubsection{Supernovae}

Some supernovae have brightness temperatures high enough for their
continuum emission to be detected with VLBI arrays. Almost all of
these are core-collapse supernovae of Types II and Ib/c. They give
astronomers the unique opportunity to observe the geometric evolution
of supernovae, and some remarkable observations have been carried out
in the last few years.

The most prominent supernova that has been studied with VLBI probably
is SN\,1993J in the nearby galaxy M\,81. \cite{Marcaide1993}
discovered early on that the supernova could be detected on
intercontinental baselines and began a monitoring
campaign. \cite{Marcaide1995a} found that the supernova had a
shell-like structure and \cite{Marcaide1995b} found the shell to
expand symmetrically, indicating that the shell was expanding into a
homogeneous medium. Hints of deceleration were reported in 1997 by
\cite{Marcaide1997} and were confirmed by
\cite{Bartel2000}, who had monitored SN1993J at regular intervals
with the VLBA since the explosion. \cite{Bartel2000} also found that
the deceleration of the expansion could not be explained with the
``standard'' model by \cite{Chevalier1982}. They inferred from the
progress of deceleration that SN1993J had left the phase of free
expansion and had entered a phase of adiabatic expansion, which is
dominated by the swept-up circumstellar material.

VLBI observations have recently also been used to search for
supernovae. Though not all attempts have been fruitful
(\citealt{Ulvestad2007}), sometimes interesting serendipitous
discoveries are made (\citealt{Neff2004}). In star-forming galaxies,
the densities of massive stars, and consequently the rate of
supernovae, is so high that explosions can be observed on time scales
of years. \cite{Lonsdale2006} have observed the prototypical
ultra-luminous infrared galaxy Arp\,220 with a very sensitive global
VLBI array, and discovered 49 point-like sources which they identified
as predominantly young supernovae. Comparing their observations to
images made 12 months earlier they were able to identify four
supernovae which had exploded in the time between the
observations. Comparing their findings to the luminosities of
supernovae in the nearby star-forming galaxy M\,82, they confirm that
the star formation in Arp\,220 is like that in M\,82 but scaled up by
a factor of a few tens.

\subsubsection{X-ray binaries}

X-ray binaries (XRBs) are thought to consist of a compact star (a
neutron star or a black hole), and a donor star from which matter is
transferred onto the compact companion. Some XRBs exhibit radio
emission and, because the accretion and emission mechanisms are
thought to be similar to those in quasars, XRBs are sometimes called
microquasars.

The properties of XRBs, when compared to quasars, scale very roughly
with luminosity, and this is also true for the time scales on which
they evolve. XRBs are variable on time scales of days, or even
hours. \cite{Fomalont2001} demonstrated this very clearly on Sco\,X-1,
observing the object continuously for more than two days, using three
VLBI arrays\footnote{see also
http://www.nrao.edu/pr/2001/scox1/scox100.htm}. Their images show a
two-sided structure with radio-emitting components moving away from a
central radio component as the observation progressed. The components
ejected from the nucleus travelled at approximately half the speed of
light. VLBI observations of the microquasar GRS\,1915+105 by
\cite{Dhawan2000} had already found speeds in excess of the speed of
light, which arose from a projection effect and is common in
observations of AGN jets. Similarly, \cite{Miller-Jones2004} find that
Cyg\,X-1 displays a two-sided jet morphology with mildly relativistic
speeds. Another instructive movie, made from daily VLBA 1.5\,GHz
snapshots over the course of six weeks, shows the microquasar
SS\,433\footnote{http://www.nrao.edu/pr/2004/ss433}
(\citealt{Mioduszewski2004}). The observations showed that the jet
ejecta brighten at specific distances from the nucleus, but only at
certain times in the 164-day precession period of the accretion disk,
indicating a non-symmetric obstacle with which the jet collides.

VLBI observations have recently helped to determine the nature of the
XRB LS\,I\,+61\,303. It had been debated whether the emission is
microquasar-like or due to a pulsar wind nebula (an illustration of
these competing models can be found in \citealt{Mirabel2006}). The
morphology of the mas-scale radio emission, monitored over one orbital
period of the binary by \cite{Dhawan2006}, indicates that the radio
emission is due to a pulsar wind nebula; the emission has a cometary
appearance and points away from the high-mass companion throughout the
orbit of the neutron star.


\subsection{Results from astrometric observations}

VLBI observations not only provide extremely high resolution but can
also yield very accurate measurements of the positions of radio
sources. In this section we describe recent results from VLBI
observations where the positional accuracy was essential.

\subsubsection{Reference frames}

Assigning a coordinate to a celestial object is not trivial. The earth
rotates, precesses and nutates, its surface changes via tectonic plate
motions, its angular velocity decelerates as a result of tidal
friction due to the moon's orbit, and the solar system rotates around
the Galactic centre at a speed of 220\,km\,s$^{-1}$. Furthermore,
planets, stars and other celestial bodies have proper motions and
change their appearances and shapes, even on time scales of years or
less. Hence it is important to have a coordinate system (a reference
system) that is tied to celestial phenomena that change as little as
possible (a reference frame, which is an implementation of the
reference system), and which can be defined as accurately as
possible. The current reference frame adopted by the International
Astronomical Union is the International Celestial Reference Frame, or
ICRF (\citealt{Ma1998}), which is based on VLBI observations of
point-like, strong, distant quasars. This reference frame satisfies
the two aforementioned requirements: quasars are very distant and are
not connected with one another in any way, hence no rotation is
expected, and their positions can be observed with very high
accuracy. Previous reference frames such as the ``Fifth Fundamental
Catalogue'' (\citealt{Fricke1988}) had been established by
observations of several thousand stars and had accuracies of the order
of tens of mas. Stars, however, have detectable proper motions and are
gravitationally bound in the Milky Way, and the accuracy of their
position measurements are much lower than those of typical VLBI
observations. The 212 ``defining'' sources in the ICRF have typical
position errors of 0.4\,mas, and the orientation of the axis defined
from this set of sources has an accuracy of 0.02\,mas. A complication
with using VLBI observations of quasars though is that radio jets
structures are subject to changes, which needs to be taken into
account for these measurements. A good review about reference systems,
reference frames and the implementation of the ICRF can be found in
\cite{Feissel1998}. Another application of astrometric observations is to
calibrate Global Positioning System (GPS) measurements. Some VLBI
stations are outfitted with GPS receivers, the positions of which are
determined using astrometric observations. The same GPS receivers are
then used to determine errors in the orbits of GPS satellites.

\cite{Lestrade1999} have carried out VLBI observations of 11
radio-emitting stars to tie the Hipparcos reference frame to the
ICRF. In addition to the Hipparcos link, these observations also had
astrophysical implications for the observed stars. For example, they
found that the binary UX\,Ari exhibits significant acceleration, which
could be caused by a yet unknown companion.

\subsubsection{Distances and proper motions of stars and masers in the
Milky Way}

Stars can be observed only if they emit non-thermal radiation, which
is typically the case for T\,Tauri stars. \cite{Torres2007} report on
VLBI observations of Hubble\,4 and HDE\,283572, two T\,Tauri stars in
the Taurus association. They infer distances of $(132.8\pm0.5)$\,pc
and $(128.5\pm0.6)$\,pc, respectively, which is an order of magnitude
more accurate than other distance measurements. In a companion paper,
\cite{Loinard2007} report a similar measurement for T\,Tauri itself. 

\cite{Sandstrom2007} present a similar measurement for the star GMR\,A
in the Orion Nebula Cluster. They find a distance of
$(389^{+24}_{-21})$\,pc, in disagreement with the hitherto canonical
value of $(480\pm80)$\,pc reported by
\cite{Genzel1981}. Placing the Orion cluster at a distance 20\,\%
smaller than previously thought implies that the luminosities of stars
stars within it are lower by a factor of 1.5 -- and that the stars are
almost twice as old, since age scales as luminosity to the $-3/2$
power.

As was mentioned in the overview of the world's VLBI arrays, Japan has
built a four-station network specifically for precise astrometry of
Galactic masers, and first results are beginning to be published:
\cite{Hirota2007} report VLBI observations of the water maser emission
from Orion KL over two years, from which they derive a trigonometric
distance of $(437\pm19)$\,pc. This measurement is difficult to
reconcile with the result reported by
\cite{Sandstrom2007} - which is the correct value time will show.
In two other papers, \cite{Hirota2007b} report a similar measurement of
the distance to the young stellar object SVS\,13 in NGC\,1333, and
\cite{Honma2007} have measured the distance to the star-forming region
Sharpless 269, finding a value of $(5.28^{+0.24}_{-0.22})$\,kpc (or a
parallax of $189\pm8\,\mu$as), which they claim is the smallest
parallax ever measured (and indeed {\it is} more precise than the
parallax of Sco\,X-1 measured by \citealt{Bradshaw1999}, who obtained
(360$\pm$40)\,$\mu$as and had once claimed the record).

Focussing on distances to evolved stars, \cite{Vlemmings2007} have
observed the OH maser transition at 1.6\,GHz in the envelopes of
S\,CrB, RR\,Aql, and U\,Her. Their measurements have accuracies of
4\,\% to 34\,\%, which is a vast improvement over older Hipparcos
measurements, with errors of 50\,\% to 100\,\%.

\subsubsection{Distances and proper motions of pulsars}

The only model-independent method to determine the distance to a
celestial object is the trigonometric parallax. From optical
observations, parallaxes can be obtained only for close
objects. Remember that the positional accuracy of optical observations
is limited to about 0.1'' and therefore parallaxes can be measured for
objects within only 10\,pc. VLBI observations, however, can yield
positions as good as 0.1\,mas and so one can measure parallaxes of
objects at a distance of up to 10\,kpc. Pulsars are among those few
objects that have brightness temperatures high enough for VLBI
observations and it is known that they are point sources (this is even
true for VLBI observations). Parallax measurements can therefore yield
distances to pulsars in large portions of the Milky Way and help to
calibrate the distance scale derived from the pulse dispersion
measurements obtained from single-dish observations.

Only a handful of pulsar parallaxes had been measured until 2000 (for
a list and references see Table~1 in \citealt{Brisken2002}). This has
changed in the last few years, and many more measurements are now
becoming available (e.g., \citealt{Brisken2002}, \citealt{Dodson2003},
\citealt{Chatterjee2005}, and \citealt{Helfand2007}). These
observations not only added to the number of trigonometric parallaxes,
but also yielded pulsar proper motions (transverse to the line of
sight). Proper motion measurements allow one to determine the likely
birth places, as done by \cite{Ng2007}, help to constrain velocities
derived from pulsar timing, or, in some cases, are the only available
velocity measurements when pulsar timing is not possible (e.g.,
\citealt{Dodson2003}).

\begin{figure}
\includegraphics[width=0.45\linewidth]{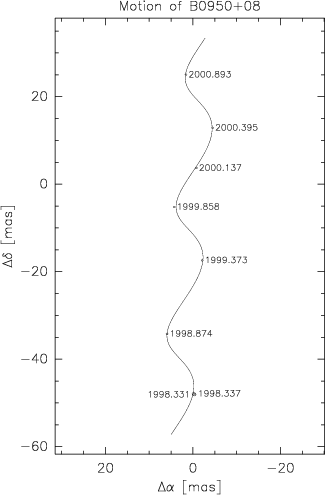}
\includegraphics[width=0.45\linewidth]{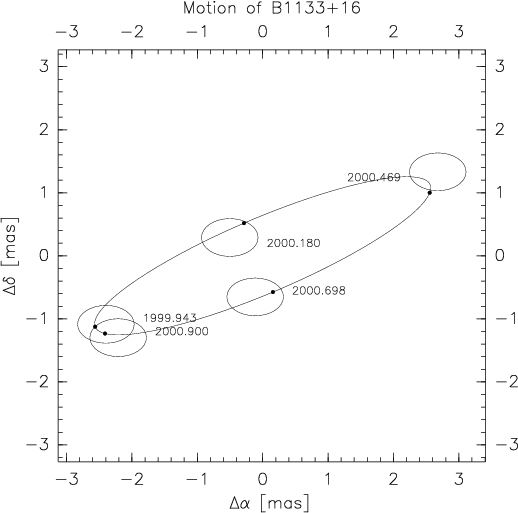}
\caption{Illustration of the combination of proper motion and parallax
motion of pulsars observed with VLBI (\citealt{Brisken2002}). The left
panel shows the observed locations of the pulsar B0950+08 over the
course of 2.5\,yr. The track is a combination of the pulsar's proper
motion and the earth's rotation around the sun. To indicate the
significance of such parallax measurements, the modelled proper motion
of the pulsar B1133+18 has been subtracted in the right panel, clearly
showing the annual parallax (the small ellipses indicate the position
errors). Reproduced with kind permission of the author.}
\label{fig:brisken}
\end{figure}

\subsubsection{Proper motions of galaxies in the Local Group}

Measurements of the proper motions in the Local Group certainly is not
a widely spread application of VLBI observations, but it illustrates
the potential of high-precision position measurements.

\begin{figure}
\resizebox{0.7\hsize}{!}{\includegraphics[bbllx=2.5cm,bburx=17.5cm,bblly=10.5cm,bbury=27cm,clip=,angle=0]{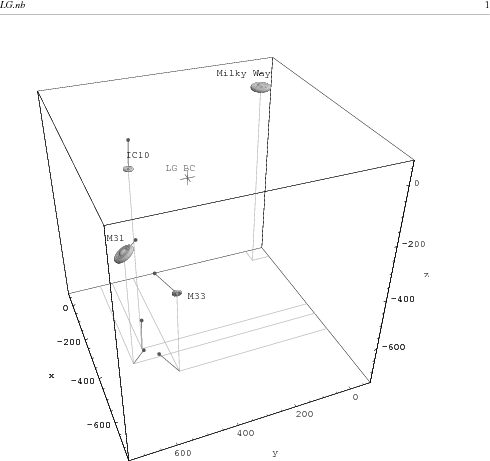}}
\resizebox{0.7\hsize}{!}{\includegraphics[bbllx=2.5cm,bburx=17.5cm,bblly=10.5cm,bbury=27cm,clip=,angle=0]{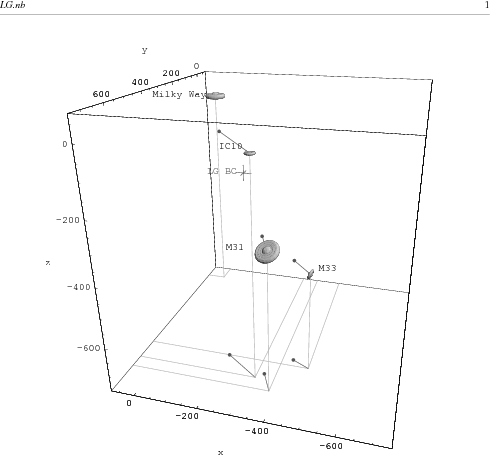}} 
\caption{The relative positions and velocities of galaxies in the Local
Group as determined by \cite{Brunthaler2005a}. Reproduced with kind
permission of the author.}
\label{fig:brunthaler}
\end{figure}

\cite{Brunthaler2005a} used the VLBA to monitor water masers
in M\,33 over more than 4\,yr. They were able to detect proper motion
of 36\,\uas\,yr$^{-1}$ in right ascension and 13\,\uas\,yr$^{-1}$ in
declination. Comparing these measurements to rotation models of M\,33
based on measurements of neutral hydrogen, they were able to infer a
geometric distance to M\,33 of (730$\pm$168)\,kpc.  Together with
similar measurements in the nearby galaxy IC\,10
(\citealt{Brunthaler2007}), a picture of the dynamics in the Local
Group begins to emerge (Fig.~\ref{fig:brunthaler} and
\citealt{Loeb2005}).


\subsubsection{Gravitational lenses}

Gravitational lenses in which the emission of a distant quasar is
deflected by a foreground galaxy, producing multiple images extended
over a few arcseconds, are obvious VLBI targets. High-resolution
observations yield accurate separation measurements and can confirm
that images belong to a lens because they have the same spectral
indices, surface brightness, or degree of polarization. If the lensed
object has sub-structure such as jets, then it becomes possible to
investigate sub-structures in the lens galaxy.

For example, \cite{Bradac2002} have observed the lens system B1422+231
and are able to produce a reasonable model for the lens only when
substructure is included (that is, they were unable to model the lens
with a simple ellipsoid). Substructure in the lens can also be
inferred from asymmetries in the lensed images, as has been
tentatively found in B1152+199 by \cite{Rusin2002}. \cite{Biggs2002}
report VLBI observations of B0128+437, and they speculate that the
lens could have a halo of intermediate-mass black holes
(``milli-lensing''). Finding substructure in the haloes of lensing
galaxies may yield clues about large-scale structure formation, driven
by a lack of halo dwarf galaxies in the Local Group.

\cite{Koopmans2002} and \cite{Porcas2002b} present VLBI observations of
MG\,2016+112, in which one image of the background source is extremely
stretched. This is interpreted as the radio jet of the lensed quasar
crossing the caustic of the lens, leading to extreme
magnification. \cite{Koopmans2002} speculate that with such high
magnification one might be able to observe hyper-luminal motion of the
order of (10$^2$ to 10$^3$)\,c.

Gravitational lenses offer the exciting opportunity to obtain a value
for $H_0$, using relatively simple astrophysics
(\citealt{Refsdal1964}). However, to constrain $H_0$ requires one to
use a model of the mass distribution in the lensing object, which is
mostly difficult to obtain (see \citealt{Schechter2005} for a review
of the associated difficulties). Nevertheless, \cite{York2005} present
an estimate from observations of CLASS B0218+357 with radio
interferometers on all scales and Hubble Space Telescope
observations. They derive values of
(70$\pm$5)\,km\,s$^{-1}$\,Mpc$^{-1}$ and
(61$\pm$7)\,km\,s$^{-1}$\,Mpc$^{-1}$, depending on the treatment of
the lensing galaxy.

We refer the reader to the reviews by \cite{Porcas2004} and
\cite{Biggs2005} for more information about gravitational lenses and
VLBI.

\subsubsection{Measurement of the speed of gravity waves}

General relativity predicts that light is deflected by matter,
something which had been measured for the first time by
\cite{Dyson1920}, who were able to measure the deflection of a star
near the sun during a solar eclipse. In 2002 \cite{Fomalont2003} have
carried out an experiment to measure the deflection of the light of a
quasar as Jupiter passed within 3.7', by monitoring the position of
the quasar relative to two other quasars over eight days with an
accuracy of less than 10\,$\mu$as. They were able not only to measure
the effects of Jupiter's mass, but also those of Jupiter's
velocity. The observed deflection matched the predictions from General
Relativity, but a controversy emerged over whether the result depends
on the speed of gravity waves, $c_g$, or the speed of light, $c$
(\citealt{Carlip2004, Kopeikin2005, Carlip2005}).

\subsubsection{Spacecraft tracking}

In 2005, a global VLBI array has been used to carry out a unique
experiment. The Cassini spacecraft, exploring Saturn, launched a probe
called Huygens onto Saturn's moon Titan. The goal was to measure
parameters like temperature, pressure, and wind speeds as the probe
would descent towards Titan's surface, hanging from a parachute. The
winds were to be measured using Doppler measurements in the radio
signal from Huygens to Cassini (the so called Doppler Wind Experiment,
DWE), the only communication link from the Titan probe.
Unfortunately, one of Cassini's receivers failed, resulting in the
potential loss of the DWE data transmitted by Huygens. The solution of
this problem was to measure Huygens' path and horizontal velocity in
the atmosphere by tracking, or rather eavesdropping on, its uplink to
Cassini with a global VLBI array. The Huygens radio link was designed
to operate between Huygens and Cassini, some 100\,000\,km apart, not
between Huygens and Earth at 1.2\,billion kilometres. Thus, the
Huygens signal at Earth was very weak (measured in tens to hundreds of
photons per telescope per second). In addition, most of the Huygens
descent took place while Saturn was best visible from the Pacific
ocean -- an area sparsely populated by radio telescopes. The
experiment succeeded nevertheless, resulting in extraordinarily
accurate data on the descent trajectory (with 1\,km accuracy) and
horizontal velocity. The observers were even able to detect the
swinging of Huygens below its parachute, with an amplitude of 0.6\,m
(Gurvits, priv. comm.).

\subsubsection{Motion of the sun around the Galactic centre}


The nature of the compact radio source in the Galactic Centre,
Sgr\,A$^{\rm *}$, has been the target of many VLBI observations (see
Section~\ref{sec:sgra}).  Evidence from astrometric observations of
the central mass in Sgr\,A$^{\rm *}$ was presented by
\cite{Reid1999}. They monitored the position of Sgr\,A$^{\rm *}$ over
two years with a VLBI array, and found that its apparent motion in the
Galactic plane is (5.90$\pm$0.4)\,mas\,yr$^{-1}$. This motion of
course is caused by the sun's orbit around the Galactic centre. When
the sun's orbit (known from other observations) was taken into
account, the proper motion of Sgr\,A$^{\rm *}$ is less than
20\,km\,s$^{-1}$. This means that Sgr\,A$^{\rm *}$ is extremely close
to the dynamic centre of the galaxy. Were the radio emission from
Sgr\,A$^{\rm *}$ emitted by an exotic form of stellar cluster, then
much higher velocities would be expected. Taking into account the mass
of the stars orbiting Sgr\,A$^{\rm *}$, which is of the order of
10\,$M_\odot$, and their high velocities of a few hundred kilometres
per second, the negligible proper motion of Sgr\,A$^{\rm *}$ implies
that it must have a mass of at least 1000\,M$_\odot$. All evidence is
consistent with Sgr\,A$^{\rm *}$ being a black hole.

\section{Conclusions}

In the general astronomical community VLBI observations are often
considered to be rather exotic, because the technique is regarded as
complicated and its range of application limited to AGN. We
acknowledge that VLBI observations, and in particular the data
calibration, can be a challenge, but we have shown here that the
technique is applicable to a surprisingly broad variety of
astrophysical objects and problems. We wish to highlight a few key
contributions.

The structure and evolution of plasma jets ejected from compact
objects such as black holes and neutron stars would not have been
possible without VLBI. In particular multi-epoch observations have
revealed the evolution of plasma jets and yielded constraints on their
structure, physics, and environment. 

In the Milky Way a key contribution of VLBI observations are proper
motions and parallaxes for stars, masers, and pulsars. These
observations are a reference for other measurements of distances and
so are fundamental for the interpretation of other observations, and
to derive a consistent picture of the Milky Way's structure.

A more general application is the use of VLBI to implement a reference
frame, which is used throughout observational astronomy. These
observations not only tell us where objects are, but they also tell us
how fast the earth rotates, and hence how to set our clocks, and how
observations at different wavelengths fit together.

Recent advances in computer technology have had profound impact on
VLBI observations, and will do so in the near future. Directly-linked
antennas will become the standard mode of operation at many VLBI
facilities, and software correlators will allow observers to image
wider fields of view, and to use more bandwidth to increase
sensitivity and $(u,v)$ coverage. The most relevant improvement
however will be the SKA: it will be sensitive enough for observations
of thermal sources. At mas-scale resolution, the SKA will image many
astronomical objects which are inaccessible today.

\bibliography{refs}

\end{document}